\documentclass[journal]{IEEEtran}
\usepackage{cite}

\usepackage{graphicx}

\usepackage{enumitem}

\ifCLASSOPTIONcompsoc
  \usepackage[caption=false,font=normalsize,labelfont=sf,textfont=sf]{subfig}
\else
  \usepackage[caption=false,font=footnotesize]{subfig}
\fi

\ifCLASSINFOpdf
\else
\fi
\usepackage[cmex10]{amsmath}
\usepackage{amssymb}
\renewcommand{\vec}[1]{\mathrm{\textbf{#1}}}

\usepackage{blindtext}
\usepackage[linesnumbered]{algorithm2e}
\DontPrintSemicolon

\usepackage{epstopdf}

\usepackage{tikz}

  \usepackage{pgfplots}
  \pgfplotsset{compat=newest}
  \usetikzlibrary{plotmarks}
  \usepackage{grffile}
  \pgfplotsset{plot coordinates/math parser=false}
  \newlength\figureheight
  \newlength\figurewidth
\graphicspath{{graphics/}}
\DeclareGraphicsExtensions{.pdf,.jpeg,.png}

\usetikzlibrary{calc,trees,positioning,arrows,chains,shapes.geometric,%
 decorations.pathreplacing,decorations.pathmorphing,shapes,%
 matrix,shapes.symbols,plotmarks,decorations.markings,shadows,fit,patterns,through,intersections}
\usetikzlibrary{spy,backgrounds}
\usepgfplotslibrary{groupplots}

\pgfplotsset{compat=newest}
\pgfplotsset{filter discard warning=false} 
\pgfplotsset{every axis label/.append style={font=\small}}
\pgfplotsset{every tick label/.append style={font=\footnotesize}}
\pgfsetplotmarksize{2.5pt}
\tikzset{
 every picture/.style={>=latex},
 every pin/.append style={font=\tiny},
 every pin edge/.append style={shorten >=1pt, shorten <=.5pt},
}
\tikzstyle{zerosep}    =  [ inner sep=0pt ]

\tikzstyle{HYPERSTATE} =  [ fill=black!10!white, ]
\tikzstyle{SUBSTATE}   =  [ fill=black!30!white, ]
\tikzstyle{EMTY}       =  [ fill=white, ]
\tikzstyle{STATE}      =  [ fill=black!40!white, ]
\tikzstyle{INPUT}      =  [ fill=black!20!white, ]
\tikzstyle{UNCOD1}     =  [ pattern=crosshatch ]
\tikzstyle{UNCOD2}     =  [ opacity=.4,fill=yellow!50!orange ]
\tikzstyle{UNCOD3}     =  [ opacity=.4,fill=yellow!20!orange ]
\tikzstyle{UNCOD1L}    =  [ draw=yellow!80!orange, ]
\tikzstyle{UNCOD2L}    =  [ draw=yellow!50!orange ]
\tikzstyle{UNCOD3L}    =  [ draw=yellow!20!orange ]
\tikzstyle{EXTEND}     =  [ pattern=north east lines, ]
\tikzstyle{TRRSSE}     =  [ pattern=north west lines, ]
\tikzstyle{XSB}        =  [ thick, |-|, shorten <=3pt, shorten >=3pt ]

\tikzstyle{spectrum} = [
 pattern=crosshatch dots,%
 pattern color=gray,
]
\tikzset{orientation/.is choice,
    orientation/lr/.style={anchor=west,right=1},
    orientation/lr2/.style={anchor=west,right=2},
    orientation/lrd/.style={anchor=west,below=1},
    orientation/lrd2/.style={anchor=west,below=2},
    orientation/rl/.style={anchor=east,left=1},
    orientation/rl2/.style={anchor=east,left=2},
    orientation/ud/.style={anchor=north,below=1},
    orientation/du/.style={anchor=south,above=1},
    orientation/rld/.style={anchor=east,below=1},
    orientation/rld2/.style={anchor=east,below=2},
}
\tikzstyle{scare} = [
]

\tikzstyle{delayEle} = [
fill=white,
anchor=west,
rectangle,
draw=black,
minimum height=5mm,
minimum width=5mm,
inner xsep=0.5em
]

\tikzstyle{syslinear} = [
 drop shadow={shadow xshift=.6mm,
              shadow yshift=-.6mm},
 fill=white,
 anchor=west,
 rectangle,
 draw=black,
 minimum height=5mm,
 minimum width=5mm,
 inner xsep=0.5em
]

\tikzstyle{sysnonlinear} = [
 drop shadow={shadow xshift=.8mm,
              shadow yshift=-.8mm},
 fill=white,
 double,
 anchor=west,
 rectangle,
 draw=black,
 minimum height=5mm,
 minimum width=5mm,
 inner xsep=0.5em
]

\tikzstyle{syssource} = [
 anchor=west,
 ellipse,
 draw=black,
 minimum height=1.5ex,
 minimum width=1.5em,
 inner xsep=0.5em
]

\tikzstyle{syssink} = [
 anchor=west,
 ellipse,
 draw=black,
 minimum height=1.5ex,
 minimum width=1.5em,
 inner xsep=0.5em
]

\tikzstyle{syssplit} = [
 fill=black,
 draw=black,
]

\tikzstyle{sysadd} = [
 draw,circle,inner sep=-1pt,
]
\tikzstyle{sysaddmod} = [
 draw,inner sep=-1pt,rectangle,inner xsep=-.3pt,inner ysep=-.2pt
]
\tikzstyle{sysmul} = [
 draw,circle,inner sep=-1pt,
]

\definecolor{MyHSBGreen}{hsb}{0.34065,1,0.91}
\pgfplotscreateplotcyclelist{colors4}{%
 {black!100!white},
 {black!75!white},
 {black!50!white},
 {black!25!white},
}
\pgfplotscreateplotcyclelist{colors5}{%
 {black!100!white},
 {black!75!white},
 {black!50!white},
 {black!25!white},
 {densely dashed, black!100!white},
 {densely dashed, black!75!white},
}
\pgfplotscreateplotcyclelist{colors10}{%
 {black!100!white},
 {black!90!white},
 {black!80!white},
 {black!70!white},
 {black!60!white},
 {black!50!white},
 {black!40!white},
 {black!30!white},
 {black!20!white},
 {black!10!white},
}

\pgfplotscreateplotcyclelist{colorsBERMDDFSE}{%
                 every mark/.append style={fill=black},mark=o\\%
                 every mark/.append style={fill=black},mark=star\\%
                 every mark/.append style={fill=black},mark=triangle\\%
                 every mark/.append style={scale=.7,fill=black},mark=square\\%
                 every mark/.append style={fill=black},mark=diamond\\%
  densely dashed,every mark/.append style={solid},mark=pentagon*\\%
  densely dashed,every mark/.append style={solid},mark=*\\%
  densely dashed,every mark/.append style={solid},mark=triangle*\\%
  densely dashed,every mark/.append style={scale=.7,solid},mark=square*\\%
  densely dashed,every mark/.append style={solid},mark=diamond*\\%
  densely dashed,every mark/.append style={solid},mark=star*\\%
}
\pgfplotscreateplotcyclelist{colors1Empty4Full}{%
                 every mark/.append style={fill=black},mark=o\\%
  densely dashed,every mark/.append style={solid},mark=*\\%
  densely dashed,every mark/.append style={solid},mark=triangle*\\%
  densely dashed,every mark/.append style={scale=.7,solid},mark=square*\\%
  densely dashed,every mark/.append style={solid},mark=diamond*\\%
}

\pgfplotscreateplotcyclelist{colors4Empty4Full}{%
                 every mark/.append style={fill=black},mark=o\\%
                 every mark/.append style={fill=black},mark=triangle\\%
                 every mark/.append style={scale=.7,fill=black},mark=square\\%
                 every mark/.append style={fill=black},mark=diamond\\%
  densely dashed,every mark/.append style={solid},mark=*\\%
  densely dashed,every mark/.append style={solid},mark=triangle*\\%
  densely dashed,every mark/.append style={scale=.7,solid},mark=square*\\%
  densely dashed,every mark/.append style={solid},mark=diamond*\\%
}
\pgfplotscreateplotcyclelist{colors6Empty4Full}{%
                 every mark/.append style={fill=black},mark=o\\%
                 every mark/.append style={fill=black},mark=star\\%
                 every mark/.append style={fill=black},mark=triangle\\%
                 every mark/.append style={scale=.7,fill=black},mark=square\\%
                 every mark/.append style={fill=black},mark=diamond\\%
                 every mark/.append style={fill=black},mark=pentagon\\%
  densely dashed,every mark/.append style={solid},mark=*\\%
  densely dashed,every mark/.append style={solid},mark=triangle*\\%
  densely dashed,every mark/.append style={scale=.7,solid},mark=square*\\%
  densely dashed,every mark/.append style={solid},mark=diamond*\\%
}
\pgfplotscreateplotcyclelist{colors8Empty4Full}{%
                 every mark/.append style={fill=black},mark=o\\%
                 every mark/.append style={fill=black},mark=star\\%
                 every mark/.append style={fill=black},mark=triangle\\%
                 every mark/.append style={scale=.7,fill=black},mark=square\\%
                 every mark/.append style={fill=black},mark=diamond\\%
                 every mark/.append style={fill=black},mark=pentagon\\%
                 every mark/.append style={fill=black},mark=oplus\\%
                 every mark/.append style={fill=black},mark=otimes\\%
  densely dashed,every mark/.append style={solid},mark=*\\%
  densely dashed,every mark/.append style={solid},mark=triangle*\\%
  densely dashed,every mark/.append style={scale=.7,solid},mark=square*\\%
  densely dashed,every mark/.append style={solid},mark=diamond*\\%
}

\pgfplotscreateplotcyclelist{colors}{%
                 every mark/.append style={fill=black},mark=o\\%
                 every mark/.append style={fill=black},mark=star\\%
                 every mark/.append style={fill=black},mark=triangle\\%
                 every mark/.append style={scale=.7,fill=black},mark=square\\%
                 every mark/.append style={fill=black},mark=diamond\\%
                 every mark/.append style={fill=black},mark=pentagon\\%
                 every mark/.append style={fill=black},mark=oplus\\%
                 every mark/.append style={fill=black},mark=otimes\\%
                 every mark/.append style={fill=black},mark=asterisk\\%
                 every mark/.append style={fill=black},mark=+\\%
                 every mark/.append style={fill=black},mark=-\\%
                 every mark/.append style={fill=black},mark=|\\%
                 every mark/.append style={fill=black},mark=x\\%
                 every mark/.append style={fill=black},mark=*\\%
                 every mark/.append style={fill=black},mark=triangle*\\%
                 every mark/.append style={scale=.7,fill=black},mark=square*\\%
                 every mark/.append style={fill=black},mark=diamond*\\%
                 every mark/.append style={fill=black},mark=star*\\%
  densely dashed,every mark/.append style={solid,fill=gray},mark=o\\%
  densely dashed,every mark/.append style={solid,fill=gray},mark=triangle\\%
  densely dashed,every mark/.append style={scale=.7,solid,fill=gray},mark=square\\%
  densely dashed,every mark/.append style={solid,fill=gray},mark=diamond\\%
  densely dashed,every mark/.append style={solid,fill=gray},mark=star\\%
  densely dashed,every mark/.append style={solid,fill=gray},mark=pentagon*\\%
  densely dashed,every mark/.append style={solid,fill=gray},mark=*\\%
  densely dashed,every mark/.append style={solid,fill=gray},mark=triangle*\\%
  densely dashed,every mark/.append style={scale=.7,solid,fill=gray},mark=square*\\%
  densely dashed,every mark/.append style={solid,fill=gray},mark=diamond*\\%
  densely dashed,every mark/.append style={solid,fill=gray},mark=star*\\%
  mark=text,text mark=a\\%
  mark=text,text mark=b\\%
  mark=text,text mark=c\\%
  mark=text,text mark=d\\%
  mark=text,text mark=e\\%
  mark=text,text mark=f\\%
  mark=text,text mark=g\\%
}

\pgfdeclarelayer{background layer}
\pgfdeclarelayer{foreground layer}
\pgfsetlayers{background layer,main,foreground layer}

\begin{document}
%
\title{Phase Shift Keying on the Hypersphere:\\Peak Power-Efficient MIMO Communications}
%
%
%

\author{Christoph Rachinger,
        Ralf R. M\"uller
        and~Johannes B. Huber
\thanks{C. Rachinger and J. B. Huber are with the Institute for Information Transmission, University of Erlangen-Nueremberg (email: christoph.rachinger@fau.de, johannes.huber@fau.de).}
\thanks{Ralf R. M\"uller is with the Institute of Digital Communications, University of Erlangen-Nueremberg (email: ralf.r.mueller@fau.de).}
\thanks{This paper has been submitted to IEEE Transactions on Wireless Communications.}} 

\maketitle

\begin{abstract}
Phase Shift Keying on the Hypersphere (PSKH), a generalization of conventional Phase Shift Keying (PSK) for Multiple-Input Multiple-Output (MIMO)
systems, is introduced. In PSKH, constellation points are distributed on a multidimensional hypersphere. The use of such constellations with a Peak-To-Average-Sum-Power-Ratio (PASPR) of 1 allows to use load-modulated transmitters which can cope with a small backoff, which in turn results in a high power efficiency.
In this paper, we discuss several methods how to generate PSKH constellations and compare their performance. After applying conventional Pulse-Amplitude Modulation (PAM), the PASPR of the continuous time PSKH signal depends on the choice of the pulse shaping method. This choice also influences bandwidth and power efficiency of a PSKH system. In order to reduce the PASPR of the continuous transmission signal, we use spherical interpolation to generate a smooth signal over the hypersphere and present corresponding receiver techniques. Additionally, complexity reduction techniques are proposed and compared. Finally, we discuss the methods presented in this paper regarding their trade-offs with respect to PASPR, bandwidth, power efficiency and receiver complexity.
\end{abstract}




%
\IEEEpeerreviewmaketitle

\section{Introduction}
\IEEEPARstart{P}{ower} efficiency is one of the driving forces behind the development of current communication technologies.
Unfortunately, one of the main sources of power consumption are amplifiers operating at low efficiency. This holds even for state-of-the-art amplifiers. They cannot be operated at their optimal power level, because signals with suboptimal Peak-To-Average-Power-Ratio (PAPR), i.e., PAPR $> 1$, require to backoff the amplifier to avoid serious clipping. Constant envelope modulation schemes such as continuous phase modulation (CPM) provide an optimal PAPR and reduce the backoff compared to other modulation schemes, such as the widely used QAM~\cite{Amp}. While this improves the power efficiency of a transmission system, the bandwidth efficiency suffers: Since the radius in the complex plane is fixed for constant envelope transmission, phase remains the only degree of freedom to represent information. This results in a rate loss compared to QAM, which is the reason why constant envelope modulation did not receive much attention since the development of GSM-Mobile Communiation, a system employing Gaussian Minimum Shift Keying (GMSK), a variant of CPM. 

The use of MIMO systems allows for another method to increase power efficiency: Not the PAPR, but the Peak-to-Average-Sum-Power-Ratio (PASPR) of a vector-valued signal can determine the required amplifier backoff. For an arbitrary vector-valued function $\vec{x}(t) \in \mathbb{C}^n$, this quantity is defined as 
\begin{equation}
	\mathrm{PASPR}(\vec{x}(t)) = \frac{\max_t \|\vec{x}(t)\|^2}{\mathcal{E}\{\|\vec{x}(t)\|^2\}}.
\end{equation}
The PASPR is a decisive factor when recently proposed load-modulated MIMO amplifiers are used~\cite{loadMod1, loadMod2}. Since the degrees of freedom are reduced by only one for all antennas, the relative rate loss is smaller the more antennas are used~\cite{MPSK}. For massive MIMO systems, the central limit theorem (CLT) guarantees that the PASPR of the continuous-time signal becomes optimal as long as the data points are distributed on a multidimensional hypersphere. We call these constellations \emph{Phase Shift Keying on the Hypersphere} (PSKH), because it is a natural extension of ordinary PSK. In conventional MIMO with only a handful of antennas, large fluctuations of the continuous transmit signal are still possible and therefore the PASPR is far from being optimal. Thus some more adaptations are necessary in order to reduce the PASPR of the transmission signal.

The rest of this paper is organized as follows: In Sec.~\ref{sec:sysModel} we introduce our system model. Unlike in PSK, there are multiple ways to construct constellation on the hypersphere. Thus we discuss several algorithms to generate PSKH constellations and their advantages and disadvantages in Sec.~\ref{sec:constellations}. Secs.~\ref{sec:sincSq} and \ref{sec:sphInt} presents two approaches to reduce the PASPR in detail. This includes receiver structures for the corresponding signals as well as numerical results for their performance. Sec.~\ref{sec:paspr} discusses how much the previously introduced approaches reduce the PASPR of the transmitter output signal and how they affect the spectrum and thus the bandwidth efficiency. The paper ends with conclusions in Sec.~\ref{sec:conclusion}.

\section{System Model}
\label{sec:sysModel}
We define a PSKH constellations as a set of $M = 2^{R_m}$ data points $\mathcal{A} = \{\vec{a}_0, \ldots, \vec{a}_{M-1} \; | \; \vec{a}_i \in \mathbb{C}^{n_T}, \, \|\vec{a}_i\| = \sqrt E_s\}$ where $E_s$ is the energy per symbol, $n_T$ is the number of transmit antennas and $R_m$ the rate per modulation interval. Unless otherwise mentioned, these constellations are modulated using conventional PAM with a pulse shaping filter $h(t)$ with normalized symbol period $T = 1$ to generate the continuous-time transmitter output signal in the equivalent complex baseband (ECB) domain
\begin{equation}
	\label{eq:contmodel}
	\vec{s}(t) = \sum_{k = -\infty}^{\infty} \vec{x}[k] h(t - k), \quad \vec x[k] \in \mathcal{A}
\end{equation}
for a given data sequence $\langle \vec x[k] \rangle$.

If $h(t)$ is a $\sqrt{\mathrm{Nyquist}}$ filter, the channel is not frequency selective, and the corresponding matched filter is applied at the receiver, the overall model is the well known discrete-time MIMO ECB channel model
\begin{equation}
	\label{eq:discmodel}
\vec{y}[k] = \vec{Hx}[k] + \vec{n}[k]
\end{equation}
where $\vec{x}[k] \in \mathbb{C}^{n_T}, \vec{y}[k] \in \mathbb{C}^{n_R}$ are transmit and receive vector at time $k$, respectively, $\vec{H} \in \mathbb{C}^{n_T \times n_R}$ is the channel matrix and $\vec{n}[k] \in \mathbb{C}^{n_R}$ is complex i.i.d. additive white Gaussian noise with variance $\sigma^2 = \frac{N_0}{T} = N_0$ per complex component.  $n_R$ and $n_T$ denote the number of transmit and receive antennas respectively, but for the remainder of this paper we assume that $n = n_R = n_T$ and omit the time index $k$ unless confusion is possible. If other pulse shaping filters than $\sqrt{\mathrm{Nyquist}}$ are used, it will be discussed in detail.

For this work, both the continuous and the discrete time models (eqs.~\eqref{eq:contmodel} and \eqref{eq:discmodel}) are important, because the first one determines the PASPR and bandwidth of the signal whereas the latter can be used for the detection of the transmitted sequence in the receiver. 

Because every point in a PSKH constellation in $n$ dimensions has energy $E_s$ and uses $\sqrt{\mathrm{Nyquist}}$ impulses, the equivalent energy per bit for uncoded transmission is given as $E_s / R_m$ at the transmitter side. In this paper, we use two different channel models: $\vec{H}$ can be unitary, which corresponds to a vector AWGN channel after equalization. In this case, transmitter and receiver energy are equal and the received energy per bit over the one-sided noise-spectral density is given as 
\begin{equation}
	\frac{E_b}{N_0}_{\mathrm{AWGN}} = \frac{E_s}{R_m\sigma^2}.
\end{equation} 
Our second channel model is the flat-fading Rayleigh model, i.e., the entries of $\vec{H}$ are i.i.d. complex Gaussian random variables with unit variance. Thus each antenna receives an average signal energy of $E_s$ per symbol and hence the total received energy over $n$ receive antennas is $nE_s$. The \emph{average} received energy per bit over the one-sided noise-spectral density is then given as 
\begin{equation}
\frac{E_b}{N_0}_{\mathrm{Fading}} = n \frac{E_b}{N_0}_{\mathrm{AWGN}} = \frac{nE_s}{R_m\sigma^2}.
\end{equation}

We omit the subindices AWGN or Fading and instead specify the channel we use in a given scenario.


\section{PSKH Constellations}
\label{sec:constellations}

\subsection{Constellation Construction}
As explained in Sec.~\ref{sec:sysModel}, a PSKH constellation is a set of $M =  2^{R_m}$ points on the hypersphere with radius $\sqrt{E_s}$ representing $R_m$ bits. The vectors $\vec{a}_i \in \mathcal{A} \subset \mathbb{C}^n$ are $n$-dimensional corresponding to $n$ transmit antennas in a MIMO system. We note that PSKH constellations are also known as \emph{spherical codes} in literature, but to our knowledge they have never been used to improve power efficiency of communication systems by means of PASPR reduction. Without further restrictions than the radius, there are many possible ways to create constellations, which might differ quite vastly in terms of quality. 
A reasonable measure for quality, as in all PAM schemes, is the minimum distance between signal points. Optimal codes in this sense and their analytic description, however, are known only for some restricted constellation sizes and dimensions~\cite{SpheresBook}.\footnote{Some examples of such optimal packings can be found in~\cite{CodeTable}.} Therefore, we compare four different algorithms to generate PSKH constellations:
\begin{itemize}
	\item \emph{Equal Area Partitioning Algorithm} (EQPA) from~\cite{EQAlg}: Generates a constellation with equally sized areas, which are usually not the Voronoi regions of a data point.
	\item \emph{k-means Clustering} (kMC) using the spherical k-means algorithm~\cite{sphkmeans}: Generates a large number of uniformly distributed points on the sphere, clusters them using the spherical k-means algorithm.
	\item \emph{Potential Minimization} (PM): Generates particles on a sphere and minimizes the potential energy between particles. This can be done via a molecular dynamics simulation~\cite{MolDyn}.
	\item \emph{Per-Antenna PSK} (PA-PSK): Generates independent PSK constellations on each antenna, then scales them to fit the power constraint.
\end{itemize}
The algorithms can be distinguished in terms of construction complexity: EQPA and PA-PSK are analytic constructions, kMC and PM rely on numerical methods and are therefore more expensive to construct. Of course, such a construction needs to be done only once and can be computed offline. If construction is nonanalytic, it is further necessary to store the constellation in memory, which we think is reasonable to implement for $R_m \lesssim 16$. Additionally, it is possible to construct a constellation for only half the number of antennas and duplicate it, which results in a small degradation of quality.
 
In order to compare constellations with respect to their performance, we take a look at three different properties: Constellation-constrained capacities, minimum distance of the constellation (also known as \emph{packing radius} in the context of spherical codes and packings) and error probabilities.

\subsection{Capacity of PSKH}
\label{sec:capacity}

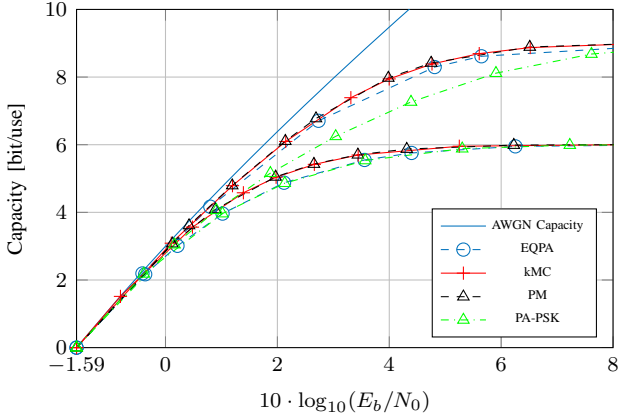
\begin{figure}[!t]
	\setlength\figureheight{4.5cm}
	\setlength\figurewidth{7.5cm} 		
%
%
\definecolor{mycolor1}{rgb}{0.00000,0.44700,0.74100}%
\definecolor{mycolor2}{rgb}{0.00000,0.44706,0.74118}%
\begin{tikzpicture}

\begin{axis}[%
width=0.951\figurewidth,
height=\figureheight,
at={(0\figurewidth,0\figureheight)},
scale only axis,
xmin=-1.59,
xmax=8,
xlabel={$10\cdot \log_{10}(E_b/N_0)$},
xmajorgrids,
xtick={-1.59, 0, 2, 4, 6, 8, 10},
ymin=0,
ymax=10,
ylabel={Capacity [bit/use]},
ymajorgrids,
axis background/.style={fill=white},
ylabel style={font=\footnotesize},xlabel style={font=\footnotesize},legend style={font=\tiny},
legend pos = {south east},
]
\addplot [color=mycolor1,solid]
  table[row sep=crcr]{%
-1.59	0\\
-0.808655789037924	1.51658088873846\\
-0.437617585816951	2.2067938175161\\
0.0647174753854372	3.11550356440512\\
0.3705819116978	3.65545808804776\\
0.714546706835895	4.25153079675009\\
1.09692462645636	4.90125769726813\\
1.5173184025604	5.60103336503701\\
1.97470392711622	6.34643165225981\\
2.46755023568392	7.13255249038584\\
2.99395570552625	7.95434510883342\\
3.55178221572521	8.80687390861694\\
4.13877427069893	9.68551176344694\\
4.75265605145154	10.5860610648317\\
};
\addlegendentry{AWGN Capacity};
\addplot [color=mycolor2,dashed, mark=o,mark options={solid}]
  table[row sep=crcr]{%
-1.59	0\\
-0.362962426475318	2.16918325203012\\
0.214060137771725	3.01019053756302\\
1.01981783379179	3.96294656612829\\
2.11920038303417	4.87618261435627\\
3.5579696347612	5.54885066020239\\
4.39883127888164	5.75594813656037\\
6.25397293326006	5.95117479960997\\
9.22024004108181	5.99757925818184\\
};
\addlegendentry{EQPA};
\addplot [color=mycolor2,dashed,mark=o,mark options={solid},forget plot]
  table[row sep=crcr]{%
-1.59	0\\
-0.411700684034853	2.19366379927365\\
0.801113511092784	4.1676251377725\\
2.73480161793833	6.70686922416324\\
4.8114037734162	8.29582577554321\\
5.64763018316539	8.61463699275132\\
9.46098757564946	8.99293061106279\\
10.4586617470588	8.99774799136607\\
};
\addplot [color=red,solid,mark=+,mark options={solid}]
  table[row sep=crcr]{%
-1.59	0\\
-0.805039985007231	1.51531875500478\\
0.481902262632733	3.56295037684444\\
1.39230171568968	4.5789914060061\\
1.97688796163821	5.03861533186732\\
2.66084670038722	5.41895232224734\\
3.44436856287592	5.69591033096007\\
5.25167005093782	5.95433129595344\\
8.2171168073656	6.00189397542168\\
10.2168527616852	6.00225889425477\\
};
\addlegendentry{kMC};
\addplot [color=red,solid,mark=+,mark options={solid},forget plot]
  table[row sep=crcr]{%
-1.59	0\\
0.10304736655805	3.0881277881713\\
1.19768245390177	4.78885574113975\\
2.14924014407941	6.09643553152199\\
3.31358816892722	7.38994460789668\\
3.99864791930418	7.94575569912032\\
4.76375508427655	8.38734469446929\\
5.60810931529684	8.69338810300434\\
6.51964881579889	8.87227753168769\\
8.46264744618111	8.98949417589356\\
10.4574191767236	9.00032272689046\\
};
\addplot [color=black,dashed,mark=triangle,mark options={solid}]
  table[row sep=crcr]{%
-1.59	0\\
0.122006117844379	3.07467621468409\\
0.889092188643982	4.08404746781762\\
1.96817334081196	5.04873604679377\\
2.65342116961028	5.42822552703081\\
3.4389993714904	5.70295655105064\\
4.31249051396057	5.87152546273956\\
6.22652316803368	5.98890857334072\\
8.21826466835946	6.00030785588569\\
10.2180706546087	6.00057591552834\\
};
\addlegendentry{PM};
\addplot [color=black,dashed,mark=triangle,mark options={solid},forget plot]
  table[row sep=crcr]{%
-1.59	0\\
0.421283628353466	3.6130305782997\\
1.19352452658776	4.79344277820347\\
2.14186244862778	6.10680081860433\\
2.69041849169461	6.7757621235787\\
3.98319590223068	7.9740767201206\\
4.74853779713467	8.41678474169889\\
6.51138470285513	8.88917650508368\\
8.46071479312624	8.99349547924501\\
10.4568817793516	9.00143649859258\\
};
\addplot [color=green,dashdotted,mark=triangle,mark options={solid}]
  table[row sep=crcr]{%
-1.59	0\\
-0.388129520967059	2.18179002620699\\
1.01889045655931	3.9637928902215\\
2.13158309163192	4.86229932693325\\
3.56967936843335	5.53390963536786\\
5.30412896021464	5.88284091111839\\
7.22702421903529	5.98821766528851\\
10.216125762849	6.00326374250044\\
};
\addlegendentry{PA-PSK};
\addplot [color=green,dashdotted,mark=triangle,mark options={solid},forget plot]
  table[row sep=crcr]{%
-1.59	0\\
0.162514419091573	3.04613087159512\\
0.91135942830053	4.06316124211033\\
1.8745974774806	5.15869975870834\\
3.04251992270395	6.24810051763609\\
4.38857593593688	7.26344088701132\\
5.90700706813934	8.11520122685276\\
7.61435915217273	8.68088663582716\\
9.49103344690882	8.93092938099495\\
10.4701520053396	8.97397384573363\\
};
\end{axis}
\end{tikzpicture}%
	\caption{Capacity of different constellations for a constellation sizes $M = 64$ and $M = 512$ points and $n = 3$ antennas over a vector AWGN channel.}
	\label{fig:capacity}
\end{figure}

In~\cite{MPSK}, it is proven that the capacity of PSKH for unitary $\vec H$ and continuous input is achieved for a uniform distribution on the hypersphere. In this section, we compare the capacities if the input is discrete and constrained to a certain size, but $\vec{H}$ is still unitary. Fig.~\ref{fig:capacity} shows the constellation constrained capacities for $n = 3$ antennas and $M = 64$ as well as $M = 512$ points. The results are similar for different constellation and antenna array sizes, which is why we restrict ourselves to one exemplary case. As a baseline, we also plot the AWGN capacity for continuous input. For $n = 3$ antennas, this capacity is $C = 3 \log(1+\mathrm{SNR_{ch}})$ with $\mathrm{SNR}_{ch} = \frac{E_s}{n \sigma^2}$ being the SNR on each individual AWGN channel. For Fig.~\ref{fig:capacity}, we use the standard representation over $\frac{E_b}{N_0}$. For the vector AWGN channel, we have $\frac{E_b}{N_0} = \frac{\mathrm{SNR_{ch}}}{C/n}$. 
The general result regarding capacities can be summarized as follows: PM and kMC have the best capacities with PM outperforming kMC by only approx. 0.01 dB. For a constellation size of 1 bit per real dimension, PA-PSK and EQPA lose up to 0.5 dB at a capacity of $C = 5.5$ bit/use. If we increase the constellation size, PM and kMC still remain on top, EQPA reduces the loss to about 0.3 dB, whereas PA-PSK loses up to 2 dB compared to PM at $C = 8.5$ bit/use.
The reason for this big loss when increasing the constellation size is that PA-PSK is the only constellation where no form of global optimization, i.e., over all 6 dimensions, takes place. While a PSK constellation might be perfect on an individual antenna, increasing the total constellation size requires to use different constellations on each antenna. This can have devastating influence on the overall distance properties of the constellation. On the other hand, EQPA works in such a way that the distribution of points becomes more and more uniform as the constellation size increases. This algorithm profits from packing the hypersphere more densely.

Fig.~\ref{fig:capacity} also shows that for coded transmission with a target rate well below $R_m$, the larger constellation can get very close to the AWGN capacity: Using $M = 512$ points per constellation and assuming a code of rate $R_c = \frac 1 2$, i.e., a total rate of $4.5$ bit, the gap to AWGN capacity is only approx. 0.1 dB. The capacities fit nicely to a coded modulation rule of thumb that 0.5 bit redundancy per real dimension for coding~\cite{ungerboeck} and another 0.5 bit redundancy for shaping~\cite{Wachsmann} are sufficient to close most of the gap to capacity.

\begin{table}[!t]
	\renewcommand{\arraystretch}{1.3}
	\caption{Average and total minimum distance of constellations}
	\label{tbl:distances}
	\centering
	\begin{tabular}{l||c|c}
		\hline
		& $d_{\mathrm{min}}$ & $d_{\mathrm{nb, avg}}$ \\
		\hline\hline
		EQPA, $M = 64$ 		& 0.6611 & 0.7282 \\
		kMC, $M = 64$ 		& 0.8674 & 0.9207 \\
		PM, $M = 64$ 		& 0.9139 & 0.9474 \\
		PA-PSK, $M = 64$ 	& 0.8165 & 0.8165 \\
		\hline
		EQPA, $M = 512$ 	& 0.4654 & 0.5350 \\
		kMC, $M = 512$ 		& 0.5235 & 0.5767 \\
		PM, $M = 512$ 		& 0.5894 & 0.6217\\
		PA-PSK, $M = 512$ 	& 0.4419 & 0.4419 \\
		\hline\hline
	\end{tabular}
	
\end{table}

\subsection{Distance Properties}
In order to compare the distance properties of the individual algorithms, Table~\ref{tbl:distances} lists the \emph{minimum distance} and \emph{average neighbor distance} of a constellation defined as
\begin{equation}
d_{\mathrm{min}}(\mathcal{A}) = \min_{\vec{a}_{i}, \vec{a}_{j} \in \mathcal{A} \atop i \neq j} ||\vec{a}_i - \vec{a}_j||
\end{equation}
and
\begin{equation}
d_{\mathrm{nb, avg}}(\mathcal{A}) = \frac 1 M \sum_{i = 0}^{M-1} \min_{\vec{a}_j \in \mathcal{A} \atop \vec{a}_i \neq \vec{a}_j} ||\vec{a}_i - \vec{a}_j||.
\end{equation}

\begin{figure*}[!t]
	\centering
	\hspace*{-1cm}
	\begin{tabular}{ll}
		\captionsetup[subfloat]{captionskip=0pt, nearskip=0.2cm, margin=0.5cm, justification=centering}
		\subfloat[Vector AWGN]{
				\setlength\figureheight{4.5cm}
				\setlength\figurewidth{7.5cm} 		
%
%
\definecolor{mycolor1}{rgb}{0.30196,0.74510,0.93333}%
\begin{tikzpicture}

\begin{axis}[%
width=0.951\figurewidth,
height=\figureheight,
at={(0\figurewidth,0\figureheight)},
scale only axis,
xmin=2,
xmax=13,
xlabel={$10 \cdot \mathrm{log}_{10}$($E_b / N_0$)},
xmajorgrids,
ymode=log,
ymin=1e-05,
ymax=1,
yminorticks=true,
ylabel={SER},
ymajorgrids,
yminorgrids,
grid style={line width=.1pt, draw=gray!10},
major grid style={line width=.2pt,draw=gray!50},
axis background/.style={fill=white},
legend style={legend cell align=left,align=left,draw=white!15!black},
ylabel style={font=\footnotesize},
xlabel style={font=\footnotesize},
legend style={font=\tiny},
]
\addplot [color=mycolor1,solid,mark=triangle,mark options={solid,rotate=180}]
  table[row sep=crcr]{%
2	0.19637896\\
2.5	0.15773112\\
3	0.12335592\\
3.5	0.09373288\\
4	0.06906016\\
4.5	0.04927912\\
5	0.03396144\\
5.5	0.02262392\\
6	0.01457552\\
6.5	0.00896472\\
7	0.00532944\\
7.5	0.0030336\\
8	0.00166776\\
8.5	0.00086856\\
9	0.000426\\
9.5	0.0001936\\
10	8.256e-05\\
10.5	3.24e-05\\
11	1.272e-05\\
11.5	3.92e-06\\
12	1.36e-06\\
};
\addlegendentry{EQPA};

\addplot [color=mycolor1,dashed,mark=triangle,mark options={solid,rotate=180}, forget plot]
  table[row sep=crcr]{%
2	0.424060903090897\\
2.5	0.366761484676207\\
3	0.310399137424035\\
3.5	0.256346402666144\\
4	0.206023720839051\\
4.5	0.160841207606352\\
5	0.121573155590407\\
5.5	0.0887099261582696\\
6	0.0622830817486767\\
6.5	0.0419451741488597\\
7	0.0270045089198196\\
7.5	0.0165916486963341\\
8	0.00969051819904594\\
8.5	0.00533882245311377\\
9	0.00276886884924525\\
9.5	0.00134215513298046\\
10	0.000606024962425668\\
10.5	0.000256028229758871\\
11	0.000104162582500163\\
11.5	3.82277984708881e-05\\
12	1.32653728027184e-05\\
12.5	4.18218649937921e-06\\
13	1.1108932888976e-06\\
13.5	3.92079984316801e-07\\
14	6.53466640528001e-08\\
14.5	0\\
15	0\\
15.5	0\\
16	0\\
16.5	0\\
17	0\\
17.5	0\\
18	0\\
18.5	0\\
19	0\\
19.5	0\\
20	0\\
};

\addplot [color=red,solid,mark=o,mark options={solid}]
  table[row sep=crcr]{%
2	0.14914408\\
2.5	0.11299952\\
3	0.0823828\\
3.5	0.05742504\\
4	0.03813456\\
4.5	0.0240384\\
5	0.01432816\\
5.5	0.00797136\\
6	0.0041664\\
6.5	0.0020052\\
7	0.00087696\\
7.5	0.00036016\\
8	0.00012864\\
8.5	3.92e-05\\
9	1.168e-05\\
9.5	3.12e-06\\
10	1.76e-06\\
10.5	0\\
11	0\\
11.5	0\\
12	0\\
};
\addlegendentry{kMC};

\addplot [color=red,dashed,mark=o,mark options={solid}, forget plot]
  table[row sep=crcr]{%
2	0.400899888910671\\
2.5	0.341751225249951\\
3	0.284172319153107\\
3.5	0.229828530353525\\
4	0.180074364503692\\
4.5	0.136177154806247\\
5	0.0990276416388943\\
5.5	0.0690555446644449\\
6	0.0458630987388094\\
6.5	0.0290079069463504\\
7	0.0173496046526825\\
7.5	0.00974658563680324\\
8	0.00514872900738417\\
8.5	0.00255263673789453\\
9	0.00117179637979481\\
9.5	0.000497810886754231\\
10	0.000194733058877344\\
10.5	6.95288505521793e-05\\
11	2.17604391295824e-05\\
11.5	6.79605306149121e-06\\
12	1.6990132653728e-06\\
12.5	3.26733320264001e-07\\
13	0\\
13.5	0\\
14	0\\
14.5	0\\
15	0\\
15.5	0\\
16	0\\
16.5	0\\
17	0\\
17.5	0\\
18	0\\
18.5	0\\
19	0\\
19.5	0\\
20	0\\
};

\addplot [color=green,solid,mark=+,mark options={solid}]
  table[row sep=crcr]{%
3	0.0800212379693162\\
4	0.0365558115616586\\
5	0.013429708175167\\
6	0.00373353795822027\\
7	0.000754293652087368\\
8	0.000102072449464669\\
9	7.81855463386609e-06\\
10	2.43743909449063e-07\\
11	0\\
12	0\\
13	0\\
};
\addlegendentry{PM};

\addplot [color=green,dashed,mark=+,mark options={solid}, forget plot]
  table[row sep=crcr]{%
3	0.275741115034229\\
4	0.171400710275284\\
5	0.0915646485374148\\
6	0.0403725917513905\\
7	0.0141477795829862\\
8	0.00375197139768255\\
9	0.00070450910253145\\
10	8.84212117826552e-05\\
11	7.48896486563383e-06\\
12	4.17839241430214e-07\\
13	0\\
};

\addplot [color=black,solid]
  table[row sep=crcr]{%
2	0.20521136\\
2.5	0.16541088\\
3	0.12975896\\
3.5	0.09872928\\
4	0.07269816\\
4.5	0.05166528\\
5	0.03525848\\
5.5	0.02303864\\
6	0.01434576\\
6.5	0.00843352\\
7	0.00466984\\
7.5	0.00240928\\
8	0.00114384\\
8.5	0.00049544\\
9	0.00020104\\
9.5	7.136e-05\\
10	2.048e-05\\
10.5	5.6e-06\\
11	1.52e-06\\
11.5	1.6e-07\\
12	0\\
};
\addlegendentry{PA-PSK};

\addplot [color=black,dashed]
  table[row sep=crcr]{%
2	0.557405214663791\\
2.5	0.509404495850487\\
3	0.459684833039273\\
3.5	0.408970790041168\\
4	0.358245376723518\\
4.5	0.308258250016337\\
5	0.260124746781677\\
5.5	0.214860550218911\\
6	0.173302359014572\\
6.5	0.136097758609423\\
7	0.103778997582173\\
7.5	0.076638632947788\\
8	0.0546294844148206\\
8.5	0.0374371038358492\\
9	0.0245872704698425\\
9.5	0.0153599294256028\\
10	0.00908285957001895\\
10.5	0.00506384369077959\\
11	0.00262758936156309\\
11.5	0.00127805005554466\\
12	0.00056988825720447\\
12.5	0.000228974710841012\\
13	8.03110501208913e-05\\
13.5	2.54198523165392e-05\\
14	6.92674638959681e-06\\
14.5	1.8297065934784e-06\\
15	1.960399921584e-07\\
15.5	6.53466640528001e-08\\
16	0\\
16.5	0\\
17	0\\
17.5	0\\
18	0\\
18.5	0\\
19	0\\
19.5	0\\
20	0\\
};

\end{axis}
\end{tikzpicture}%
			\label{fig:subSER1}} &
		\captionsetup[subfloat]{captionskip=0pt, nearskip=0.2cm, margin=0.5cm, justification=centering}
		\subfloat[Rayleigh Fading]{
				\setlength\figureheight{4.5cm}
				\setlength\figurewidth{7.5cm} 
			\hspace*{-.5cm}
%
%
\definecolor{mycolor1}{rgb}{0.30196,0.74510,0.93333}%
\definecolor{mycolor2}{rgb}{0.85000,0.32500,0.09800}%
\begin{tikzpicture}

\begin{axis}[%
width=0.951\figurewidth,
height=\figureheight,
at={(0\figurewidth,0\figureheight)},
scale only axis,
xmin=7,
xmax=20,
xlabel={$10 \cdot \mathrm{log}_{10}$($E_b / N_0$)},
xmajorgrids,
ymode=log,
ymin=1e-05,
ymax=1,
yminorticks=true,
ylabel={SER},
ymajorgrids,
yminorgrids,
grid style={line width=.1pt, draw=gray!10},
major grid style={line width=.2pt,draw=gray!50},
axis background/.style={fill=white},
legend style={legend cell align=left,align=left,draw=white!15!black},
ylabel style={font=\footnotesize},xlabel style={font=\footnotesize},legend style={font=\tiny},
]
\addplot [color=mycolor1,solid,mark=triangle,mark options={solid,rotate=180}]
  table[row sep=crcr]{%
0.771212547196624	0.47598024395121\\
1.77121254719662	0.398455448910218\\
2.77121254719662	0.321646870625875\\
3.77121254719662	0.24940073985203\\
4.77121254719662	0.185282143571286\\
5.77121254719662	0.131725234953009\\
6.77121254719662	0.0895679064187163\\
7.77121254719662	0.0583286142771446\\
8.77121254719662	0.0364373325334933\\
9.77121254719662	0.021944851029794\\
10.7712125471966	0.0127727854429114\\
11.7712125471966	0.00722075584883023\\
12.7712125471966	0.00397054589082184\\
13.7712125471966	0.00213881223755249\\
14.7712125471966	0.00114391121775645\\
15.7712125471966	0.000609358128374325\\
16.7712125471966	0.000327814437112577\\
17.7712125471966	0.000182843431313737\\
18.7712125471966	0.000106918616276745\\
19.7712125471966	6.53269346130774e-05\\
20.7712125471966	4.07718456308738e-05\\
21.7712125471966	2.61147770445911e-05\\
};
\addlegendentry{EQPA};

\addplot [color=mycolor1,dashed,mark=triangle,mark options={solid,rotate=180},forget plot]
  table[row sep=crcr]{%
6.77121254719662	0.235236433333333\\
7.27121254719662	0.1998455\\
7.77121254719662	0.167734666666667\\
8.27121254719662	0.139177466666667\\
8.77121254719662	0.114124633333333\\
9.27121254719662	0.0925366666666667\\
9.77121254719662	0.0741304\\
10.2712125471966	0.0587988333333333\\
10.7712125471966	0.0461295666666667\\
11.2712125471966	0.0358092666666667\\
11.7712125471966	0.0275639666666667\\
12.2712125471966	0.0210306333333333\\
12.7712125471966	0.0159048\\
13.2712125471966	0.0119278333333333\\
13.7712125471966	0.00888853333333333\\
14.2712125471966	0.0065841\\
14.7712125471966	0.0048503\\
15.2712125471966	0.00354516666666667\\
15.7712125471966	0.00258296666666667\\
16.2712125471966	0.0018719\\
16.7712125471966	0.00135543333333333\\
17.2712125471966	0.000979366666666667\\
17.7712125471966	0.000706466666666667\\
18.2712125471966	0.0005078\\
18.7712125471966	0.0003679\\
19.2712125471966	0.000266633333333333\\
19.7712125471966	0.0001926\\
20.2712125471966	0.000142966666666667\\
20.7712125471966	0.0001054\\
21.2712125471966	7.81333333333333e-05\\
21.7712125471966	5.88e-05\\
22.2712125471966	4.43333333333333e-05\\
22.7712125471966	3.38666666666667e-05\\
23.2712125471966	2.59e-05\\
23.7712125471966	1.99666666666667e-05\\
24.2712125471966	1.55666666666667e-05\\
24.7712125471966	1.23333333333333e-05\\
};
\addplot [color=red,solid,mark=o,mark options={solid}]
  table[row sep=crcr]{%
6.77121254719662	0.07591546\\
7.27121254719662	0.06076434\\
7.77121254719662	0.04812896\\
8.27121254719662	0.03774018\\
8.77121254719662	0.02929968\\
9.27121254719662	0.02254616\\
9.77121254719662	0.01719066\\
10.2712125471966	0.01300958\\
10.7712125471966	0.00975402\\
11.2712125471966	0.00727788\\
11.7712125471966	0.00539002\\
12.2712125471966	0.0039664\\
12.7712125471966	0.00290484\\
13.2712125471966	0.00212064\\
13.7712125471966	0.00153672\\
14.2712125471966	0.0011113\\
14.7712125471966	0.00080438\\
15.2712125471966	0.00058\\
15.7712125471966	0.0004202\\
16.2712125471966	0.00030012\\
16.7712125471966	0.00021432\\
17.2712125471966	0.00015358\\
17.7712125471966	0.00010864\\
18.2712125471966	7.698e-05\\
18.7712125471966	5.516e-05\\
19.2712125471966	3.874e-05\\
19.7712125471966	2.684e-05\\
20.2712125471966	1.852e-05\\
20.7712125471966	1.268e-05\\
21.2712125471966	8.66e-06\\
21.7712125471966	6.32e-06\\
22.2712125471966	4.52e-06\\
22.7712125471966	3.08e-06\\
23.2712125471966	2.16e-06\\
23.7712125471966	1.56e-06\\
24.2712125471966	1.08e-06\\
24.7712125471966	7.6e-07\\
};
\addlegendentry{kMC};

\addplot [color=green,solid,mark=+,mark options={solid}]
  table[row sep=crcr]{%
1	0.434950105010501\\
2	0.35603495349535\\
3	0.279914236423642\\
4	0.210654755475548\\
5	0.151425322532253\\
6	0.103841509150915\\
7	0.0680231923192319\\
8	0.0426500900090009\\
9	0.0256766126612661\\
10	0.0149442244224422\\
11	0.00843214821482148\\
12	0.00463459345934593\\
13	0.00249581458145815\\
14	0.00132625262526253\\
15	0.00068966396639664\\
16	0.000353375337533753\\
17	0.000179042904290429\\
18	8.97089708970897e-05\\
19	4.34293429342934e-05\\
20	2.06870687068707e-05\\
};
\addlegendentry{PM};

\addplot [color=green,dashed,mark=+,mark options={solid},forget plot]
  table[row sep=crcr]{%
-0.228787452803376	0.785168697506999\\
0.771212547196624	0.72475433942141\\
1.77121254719662	0.652798106919077\\
2.77121254719662	0.570812091721104\\
3.77121254719662	0.482014371417144\\
4.77121254719662	0.391036981735769\\
5.77121254719662	0.303544314091454\\
6.77121254719662	0.224808425543261\\
7.77121254719662	0.158620157312358\\
8.77121254719662	0.106645847220371\\
9.77121254719662	0.0684092920943874\\
10.7712125471966	0.0420173843487535\\
11.7712125471966	0.0247980002666311\\
12.7712125471966	0.0141351153179576\\
13.7712125471966	0.00781227836288495\\
14.7712125471966	0.00421662445007332\\
15.7712125471966	0.00223147580322624\\
16.7712125471966	0.00115960538594854\\
17.7712125471966	0.000591841087854953\\
18.7712125471966	0.000299320090654579\\
19.7712125471966	0.000153472870283962\\
20.7712125471966	7.62964938008266e-05\\
21.7712125471966	3.76616451139848e-05\\
};
\addplot [color=black,solid]
  table[row sep=crcr]{%
0.771212547196624	0.475902879424115\\
1.77121254719662	0.397994461107778\\
2.77121254719662	0.320768286342731\\
3.77121254719662	0.248179604079184\\
4.77121254719662	0.18383325334933\\
5.77121254719662	0.130173485302939\\
6.77121254719662	0.0881417716456709\\
7.77121254719662	0.0571179564087183\\
8.77121254719662	0.0355495700859828\\
9.77121254719662	0.0213408518296341\\
10.7712125471966	0.0124094981003799\\
11.7712125471966	0.00701485702859428\\
12.7712125471966	0.0038869426114777\\
13.7712125471966	0.00213375324935013\\
14.7712125471966	0.00116468706258748\\
15.7712125471966	0.000645110977804439\\
16.7712125471966	0.000367206558688262\\
17.7712125471966	0.000218596280743851\\
18.7712125471966	0.000137332533493301\\
19.7712125471966	9.15216956608678e-05\\
20.7712125471966	6.38472305538892e-05\\
21.7712125471966	4.54309138172366e-05\\
};
\addlegendentry{PA-PSK};

\addplot [color=black,dashed,forget plot]
  table[row sep=crcr]{%
6.77121254719662	0.310862966666667\\
7.27121254719662	0.271530766666667\\
7.77121254719662	0.234505166666667\\
8.27121254719662	0.200195366666667\\
8.77121254719662	0.168958066666667\\
9.27121254719662	0.1409549\\
9.77121254719662	0.116289833333333\\
10.2712125471966	0.0948222333333333\\
10.7712125471966	0.0764976333333333\\
11.2712125471966	0.0610140666666667\\
11.7712125471966	0.0481457\\
12.2712125471966	0.0376413666666667\\
12.7712125471966	0.0291445333333333\\
13.2712125471966	0.0223650333333333\\
13.7712125471966	0.0170174\\
14.2712125471966	0.0128475\\
14.7712125471966	0.00961563333333333\\
15.2712125471966	0.0071331\\
15.7712125471966	0.00527016666666667\\
16.2712125471966	0.00387546666666667\\
16.7712125471966	0.00283646666666667\\
17.2712125471966	0.0020642\\
17.7712125471966	0.00149653333333333\\
18.2712125471966	0.0010876\\
18.7712125471966	0.000791466666666667\\
19.2712125471966	0.000576\\
19.7712125471966	0.0004192\\
20.2712125471966	0.000311366666666667\\
20.7712125471966	0.0002314\\
21.2712125471966	0.000174466666666667\\
21.7712125471966	0.0001323\\
22.2712125471966	0.000102\\
22.7712125471966	8.02666666666667e-05\\
23.2712125471966	6.39333333333333e-05\\
23.7712125471966	5.2e-05\\
24.2712125471966	4.22e-05\\
24.7712125471966	3.47666666666667e-05\\
};
\addplot [color=mycolor2,dashed,mark=o,mark options={solid},forget plot]
  table[row sep=crcr]{%
6.77121254719662	0.226728933333333\\
7.27121254719662	0.191807666666667\\
7.77121254719662	0.160352566666667\\
8.27121254719662	0.1324511\\
8.77121254719662	0.1080794\\
9.27121254719662	0.0872234666666667\\
9.77121254719662	0.0695484666666667\\
10.2712125471966	0.0549008333333333\\
10.7712125471966	0.0428984666666667\\
11.2712125471966	0.0331841\\
11.7712125471966	0.0254394\\
12.2712125471966	0.0193329\\
12.7712125471966	0.0145687333333333\\
13.2712125471966	0.0108924666666667\\
13.7712125471966	0.0080867\\
14.2712125471966	0.00596843333333333\\
14.7712125471966	0.00438393333333333\\
15.2712125471966	0.00320496666666667\\
15.7712125471966	0.00232506666666667\\
16.2712125471966	0.00168976666666667\\
16.7712125471966	0.00121803333333333\\
17.2712125471966	0.0008763\\
17.7712125471966	0.0006265\\
18.2712125471966	0.0004474\\
18.7712125471966	0.000318933333333333\\
19.2712125471966	0.0002264\\
19.7712125471966	0.0001618\\
20.2712125471966	0.000114\\
20.7712125471966	8.10666666666667e-05\\
21.2712125471966	5.77666666666667e-05\\
21.7712125471966	4.03e-05\\
22.2712125471966	2.82666666666667e-05\\
22.7712125471966	2.00666666666667e-05\\
23.2712125471966	1.46333333333333e-05\\
23.7712125471966	1.03333333333333e-05\\
24.2712125471966	7.33333333333333e-06\\
24.7712125471966	4.63333333333333e-06\\
};
\end{axis}
\end{tikzpicture}%
			\label{fig:subSER2}} 
	\end{tabular}
	\caption{Symbol error rate (SER) for transmission of $R_m = 6$ (solid) or $R_m = 9$ (dashed) bits per constellation point over a vector AWGN channel (a), and a Rayleigh fading channel (b). The system has $n = 3$ antennas.}
	\label{fig:constSER}
\end{figure*}
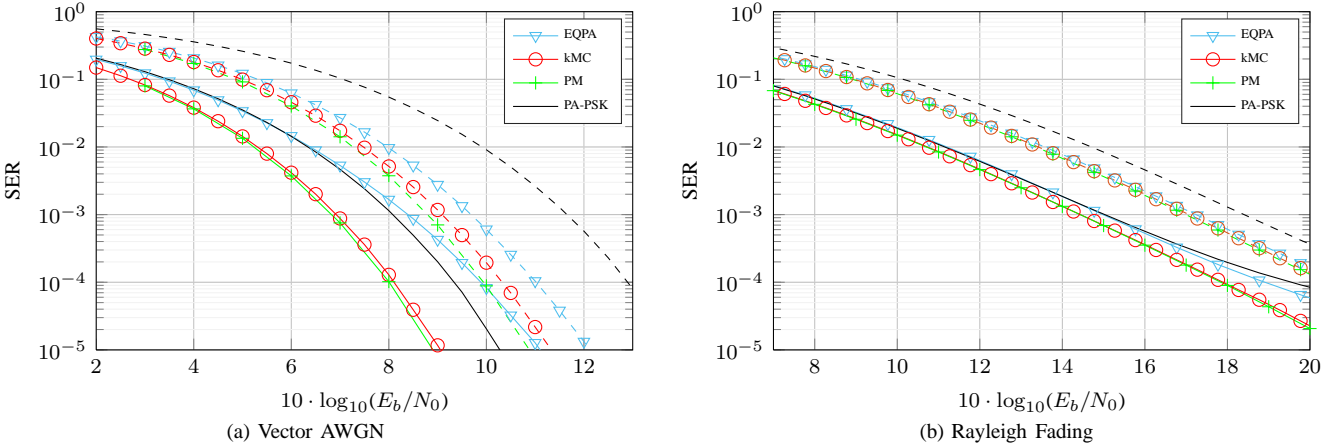

PSKH constellations, especially the ones which were generated numerically, often have asymmetric distance profiles: In conventional modulation schemes like PSK or QAM, every point has at least one neighbor which is the minimum distance apart. A lot of points (in PSK even all points) have the same distance profile to neighboring points. In PSKH, there may be only two neighboring points which are the minimum distance apart, whereas all other points in the constellation only have neighbors which are further apart. This is the reason why we also include the average neighbor distance $d_{nb, avg}$. 
The results in Table~\ref{tbl:distances} show for example, that EQPA and PA-PSK differ only slightly in terms of minimum distance for $M=512$. Nevertheless, their capacities are quite different. The qualitative result of this is that the distance profile has an effect on the capacity, but it is not possible to estimate capacities from these values alone. Various PSKH constellations with similar minimum distance might have significantly different overall distance profiles, resulting in varying capacities and performances. 

\subsection{Power Efficiency} 
In order to elaborate how the distance profile effects the power efficiency of PSKH constellations, Fig.~\ref{fig:constSER} shows the symbol error rate (SER) for constellations when 3 antennas are used with constellation sizes $M = 64$ and $M = 512$. Fig.~\ref{fig:subSER1} shows transmission over a vector AWGN channel, i.e., $\vec{H} = \vec{I}$. This corresponds to the discussion of the capacity on a channel with unitary channel matrix in Sec.~\ref{sec:capacity}. For Fig.~\ref{fig:subSER2} we use the standard Rayleigh fading model, i.e., every element of $\vec{H}$ is a complex i.i.d. Gaussian random variable with unit variance and we average over several thousand realizations. In both cases, one can observe that the error performance of the constellation is ordered according to the minimum distance of the constellation (like the capacity), but due to the size of a constellation and asymmetry of the distance profiles (see last section), a simple quantitative estimation is not developed yet. 

\emph{Remark}: We note that PA-PSK with $M = 64$ for 3 antennas is regular 4-PSK on each antenna. Without further modification, an optimized constellation such as PM can give a substantial gain compared to 4-PSK/antenna of almost 1.5 dB on the vector AWGN channel. Such a channel is equivalent to subsequent transmissions over a regular AWGN channel. Constellations which achieve a coding gain by being spread over subsequent transmissions on a Single-Input Single-Output (SISO) channel are also known as \emph{multidimensional constellations}~\cite{MultidimConst}. Multidimensional constellations can be used to combat fading~\cite{MultidimConst2, MultidimConst3}, to exploit four available dimensions in optical communications~\cite{OpticMultidimConst1, OpticMultidimConst2} and to introduce a more flexible trade-off between bandwidth the power efficiency of trellis-coded signals~\cite{CodedMultiConst}. Such constellations are usually based on conventional 2-D modulation schemes or on some lattices, which generally does not result in fixed radius constellations.  To our knowledge, no work has dealt with multidimensional constellations with fixed radius in MIMO systems to exploit load-modulation transmitters.

\section{Sinc$^2$ Pulse Shaping}
\label{sec:sincSq}

\begin{figure}[t]
	\setlength\figureheight{4.5cm}
	\setlength\figurewidth{7.5cm} 
%
%
\begin{tikzpicture}

\begin{axis}[%
width=0.951\figurewidth,
height=\figureheight,
at={(0\figurewidth,0\figureheight)},
scale only axis,
xmin=5,
xmax=20,
xlabel style={font=\color{white!15!black}},
xlabel={$10 \cdot \log_{10}(E_b/N_0)$},
ymode=log,
ymin=1e-06,
ymax=1,
yminorticks=true,
ylabel style={font=\color{white!15!black}},
ylabel={SER},
axis background/.style={fill=white},
xmajorgrids,
ymajorgrids,
yminorgrids,
legend style={at={(0.03,0.03)}, anchor=south west, legend cell align=left, align=left, draw=white!15!black},
ylabel style={font=\footnotesize},xlabel style={font=\footnotesize},legend style={font=\tiny},
]
\addplot [color=red, line width=.70pt, mark=triangle, mark options={solid, rotate=180, red}]
  table[row sep=crcr]{%
-4.98970004336019	0.669436741767764\\
-2.98970004336019	0.579219370750567\\
-0.989700043360188	0.470592800959872\\
1.01029995663981	0.352190814558059\\
3.01029995663981	0.239099186775097\\
5.01029995663981	0.146112065057992\\
7.01029995663981	0.0806206505799227\\
9.01029995663981	0.0406765497933609\\
11.0102999566398	0.0191661245167311\\
13.0102999566398	0.0086091454472737\\
15.0102999566398	0.00376097853619517\\
17.0102999566398	0.00161137181709105\\
19.0102999566398	0.000689028129582722\\
21.0102999566398	0.000297733635515265\\
23.0102999566398	0.000127982935608586\\
};
\addlegendentry{ISI-free}

\addplot [color=red, dashed, line width=.70pt, mark=triangle, mark options={solid, rotate=180, red}, forget plot]
  table[row sep=crcr]{%
-3.22878745280338	0.721007958938808\\
-1.22878745280338	0.601410038661512\\
0.771212547196624	0.453156312491668\\
2.77121254719662	0.297072310358619\\
4.77121254719662	0.164149073456872\\
6.77121254719662	0.0753689374750033\\
8.77121254719662	0.0289929742700973\\
10.7712125471966	0.00963019597387015\\
12.7712125471966	0.00287432342354353\\
14.7712125471966	0.000793534195440608\\
16.7712125471966	0.000207225703239568\\
18.7712125471966	5.22730302626316e-05\\
20.7712125471966	1.2571657112385e-05\\
22.7712125471966	3.01293160911878e-06\\
24.7712125471966	8.26556459138781e-07\\
};

\addplot [color=green, mark=triangle, mark options={solid, green}]
  table[row sep=crcr]{%
-2.98970004336019	0.587524730035995\\
-0.989700043360188	0.480208878816158\\
1.01029995663981	0.361990401279829\\
3.01029995663981	0.247515157978936\\
5.01029995663981	0.152141981069191\\
7.01029995663981	0.0843375416611118\\
9.01029995663981	0.0427018130915878\\
11.0102999566398	0.0201558058925477\\
13.0102999566398	0.00907828289561392\\
15.0102999566398	0.00396503132915611\\
17.0102999566398	0.00170669244100787\\
19.0102999566398	0.000728582855619251\\
21.0102999566398	0.000314958005599253\\
23.0102999566398	0.000136341821090521\\
};
\addlegendentry{$\mathrm{sinc}^2$, DFE}

\addplot [color=blue]
  table[row sep=crcr]{%
-2.98970004336019	0.585599146780429\\
-0.989700043360188	0.477955565924543\\
1.01029995663981	0.359469457405679\\
3.01029995663981	0.244862698306892\\
5.01029995663981	0.149821943740835\\
7.01029995663981	0.0826564991334489\\
9.01029995663981	0.0416379016131182\\
11.0102999566398	0.0195757365684575\\
13.0102999566398	0.00878198906812425\\
15.0102999566398	0.00383096920410612\\
17.0102999566398	0.00164592720970537\\
19.0102999566398	0.000703946140514598\\
21.0102999566398	0.000303346220503933\\
23.0102999566398	0.000130649246767098\\
};
\addlegendentry{$\mathrm{sinc}^2$, VA}

\addplot [color=green, dashed, mark=triangle, mark options={solid, green}]
  table[row sep=crcr]{%
-1.22878745280338	0.613327463004933\\
0.771212547196624	0.466831302493001\\
2.77121254719662	0.309505745900547\\
4.77121254719662	0.172750779896014\\
6.77121254719662	0.0798798693507532\\
8.77121254719662	0.0308808158912145\\
10.7712125471966	0.0103013064924677\\
12.7712125471966	0.00306496467137715\\
14.7712125471966	0.000846927076389815\\
16.7712125471966	0.000221583788828156\\
18.7712125471966	5.5419277429676e-05\\
20.7712125471966	1.28649513398214e-05\\
22.7712125471966	2.9729369417411e-06\\
24.7712125471966	7.9989334755366e-07\\
};

\addplot [color=blue, dashed]
  table[row sep=crcr]{%
-1.22878745280338	0.611169350753233\\
0.771212547196624	0.463966737768298\\
2.77121254719662	0.305964324756699\\
4.77121254719662	0.1692887215038\\
6.77121254719662	0.0774365284628716\\
8.77121254719662	0.0296142914278096\\
10.7712125471966	0.00978305559258765\\
12.7712125471966	0.002888881482469\\
14.7712125471966	0.000799973336888415\\
16.7712125471966	0.000207772297027063\\
18.7712125471966	5.18597520330623e-05\\
20.7712125471966	1.20783895480603e-05\\
22.7712125471966	2.71963738168244e-06\\
24.7712125471966	7.46567124383416e-07\\
};

\end{axis}
\end{tikzpicture}%
	\caption{Comparison of a ISI-free tranmission using a $\sqrt{\mathrm{Nyquist}}$ impulse and $\mathrm{sinc}^2$ pulse shaping. Transmission is over $n = 2$ (solid) and $n = 3$ (dashed) antennas with one bit per real dimension ($M = 16$ and $M = 64$). The VA uses $\nu = 2$ memory elements. Simulations were performed over several thousand Rayleigh fading channels and averaged.}
	\label{fig:sincSq}
\end{figure}
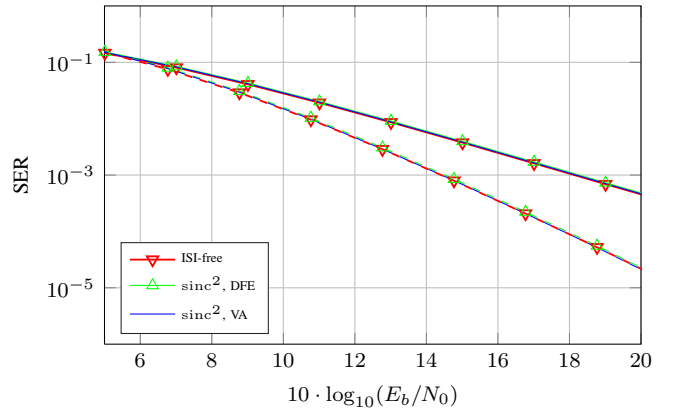

The simplest method to reduce the PASPR is to use pulse shaping filters which show better PASPR properties. PAM employing a  $\mathrm{sinc}^2(t)$-function\footnote{We define $\mathrm{sinc}(x) = \sin(\pi x) / (\pi x)$ for $x \neq 0$ and 1 otherwise.} for pulse shaping shows very good properties even for very few antennas (see Sec.~\ref{sec:paspr}). This means that the continuous transmission signal is
\begin{equation}
	\vec{s}(t) = \sum_{k = -\infty}^{\infty} \vec{x}[k] \, \mathrm{sinc}^2\left(\frac{t-kT}{T}\right).
\end{equation}
Since $\mathrm{sinc}^2$ is not a $\sqrt{\mathrm{Nyquist}}$-function, some ISI has to be equalized at the receiver. This ISI is not generated by the channel, but only by the pulse shaping filter and its corresponding matched filter, i.e., there is no ISI between different receive antennas. Thus there is no need to make use of equalization techniques developed for MIMO ISI channels. Instead, we filter the received signal of each antenna using Forney's \emph{Whitened Matched Filter} (WMF)~\cite{WMF} to get a minimum phase impulse. ISI can then be expressed by a one-dimensional, causal minimum-phase impulse $h_W[i]$ and the resulting discrete time transmission model becomes
\begin{equation}
	\vec{y}[k] = \vec{H}\sum_{i = 0}^{L} h_W[i] \vec{x}[k-i] + \vec{n}[k]
\end{equation}
with ISI-length $L$. ISI can be equalized with Maximum Likelihood Sequence Estimation (MLSE) using a vector-valued Viterbi Algorithm (VA)~\cite{Viterbi}, Decision Feedback Equalization (DFE) or Delayed Decision Feedback Sequence Estimation (DDFSE)~\cite{DDFSE}, which allows a performance trade-off between DFE and MLSE. In this specific case, almost all energy of $h_W$ is stored in the very first coefficient $h_W[0]$, such that there is only a minimal loss in terms of error probability when using DFE. Results of numerical simulations can be found in Fig.~\ref{fig:sincSq}. There is virtually no loss between $\mathrm{sinc}^2$ pulse shaping with DFE (the simplest equalization method in this scenario) and ISI-free transmission by means of a $\sqrt{\mathrm{Nyquist}}$ pulse shaping in terms of power efficiency.

\section{Spherical Interpolation Signaling}
\label{sec:sphInt}

Spherical Interpolation (SI) signaling tries to smoothen the transmission signal by forcing it onto the hypersphere also in-between data samples. This is achieved by inserting interpolation points at a certain oversampling rate. The positive effect is a significantly reduced PASPR compared to conventional PAM because the signal becomes smoother and deviations from the hypersphere are reduced, especially zero-crossings. The disadvantage is ISI introduced by the interpolation points and thus an increased receiver complexity.

Before presenting two different approaches, we define spherical interpolation also known as SLERP (SphericaL intERPolation)~\cite{SLERP}: Given two points $\vec{x}_1, \vec{x}_2 \in \mathbb{R}^N$ with $||\vec{x}_1|| = ||\vec{x}_2|| = 1$ and $\cos(\theta) =  \langle \vec{x}_1, \vec{x}_2 \rangle$, for any $0 \leq \tau \leq 1$, the spherical interpolation of $\vec{x}_1$ and $\vec{x}_2$ is given as
\begin{equation}
	\vec{SI} (\vec{x}_1, \vec{x}_2, \tau) = \frac{\sin\left((1-\tau)\theta\right)}{\sin\theta}\vec{x}_1 + \frac{\sin\left(\tau\theta\right)}{\sin\theta} \vec{x}_2.
\end{equation}
\subsection{$\frac{T}{2}$-Pulse Shaping}
\label{subsec:t2rrc}
Generating a signal using spherical interpolation values in-between $T$-spaced data symbols introduces ISI. In order to simplify equalization, our first approach is to use a $\sqrt{\mathrm{Nyquist}}$-filter with respect to $T/2$, i.e., a pulse shaping filter $h(2t)$.  The corresponding matched filter is $h^\star(-2t)$. In-between data symbols, the interpolation of two adjacent points is transmitted. This way, no two transmitted symbols are opposite of the hypersphere and hence the PASPR is reduced.
The resulting output signal of the transmitter is
\begin{align}
	\vec{s}(t) &= \sum_{k = -\infty}^{\infty} \left( \vphantom{\frac 12} \vec{x}[k] \, h\left(2 (t - kT)\right) \right. \nonumber \\
			&+ \left. \vec{SI} \left(\vec{x}[k], \vec{x}[k+1], \frac 12\right) \, h\left(2 \left( t - \left(k + \frac 1 2 \right)T\right)\right) \right)
\end{align}
which is $\sqrt{\mathrm{Nyquist}}$ with respect to half the symbol rate. Filtering with the corresponding matched filter and sampling with a rate of $\frac{T}{2}$ at the receiver gives a sequence consisting of alternating data points and interpolation values. Because we chose a filter which is $\sqrt{\mathrm{Nyquist}}$ with respect to $\frac{T}{2}$, every sample at the receiver is ISI-free.

To keep this system comparable with conventional PAM, both data and interpolation symbols contain only half the original symbol energy. Therefore it is necessary to use all points at the receiver to estimate the data sequence, otherwise half of the energy would be wasted. Data estimation for $\frac{T}{2}$-pulse shaping is done via the Viterbi algorithm. Due to the interpolation, every metric in the receiver depends on current and previous received value, i.e., the VA requires exactly $\nu = 1$ memory element.

\begin{figure}[t]
	\setlength\figureheight{4.5cm}
	\setlength\figurewidth{7.5cm} 
%
%
\definecolor{mycolor1}{rgb}{0.00000,0.74902,0.74902}%
\begin{tikzpicture}

\begin{axis}[%
width=0.951\figurewidth,
height=\figureheight,
at={(0\figurewidth,0\figureheight)},
scale only axis,
xmin=5,
xmax=20,
xlabel style={font=\color{white!15!black}},
xlabel={$10\cdot \log_{10}(E_b/N_0)$},
ymode=log,
ymin=1e-06,
ymax=1,
yminorticks=true,
ylabel style={font=\color{white!15!black}},
ylabel={SER},
axis background/.style={fill=white},
xmajorgrids,
ymajorgrids,
yminorgrids,
legend style={at={(0.03,0.03)}, anchor=south west, legend cell align=left, align=left, draw=white!15!black},
ylabel style={font=\footnotesize},xlabel style={font=\footnotesize},legend style={font=\tiny},
]
\addplot [color=red]
  table[row sep=crcr]{%
-4.98970004336019	0.669436741767764\\
-2.98970004336019	0.579219370750567\\
-0.989700043360188	0.470592800959872\\
1.01029995663981	0.352190814558059\\
3.01029995663981	0.239099186775097\\
5.01029995663981	0.146112065057992\\
7.01029995663981	0.0806206505799227\\
9.01029995663981	0.0406765497933609\\
11.0102999566398	0.0191661245167311\\
13.0102999566398	0.0086091454472737\\
15.0102999566398	0.00376097853619517\\
17.0102999566398	0.00161137181709105\\
19.0102999566398	0.000689028129582722\\
21.0102999566398	0.000297733635515265\\
23.0102999566398	0.000127982935608586\\
};
\addlegendentry{ISI-free}

\addplot [color=red, dashed, forget plot]
  table[row sep=crcr]{%
-3.22878745280338	0.721007958938808\\
-1.22878745280338	0.601410038661512\\
0.771212547196624	0.453156312491668\\
2.77121254719662	0.297072310358619\\
4.77121254719662	0.164149073456872\\
6.77121254719662	0.0753689374750033\\
8.77121254719662	0.0289929742700973\\
10.7712125471966	0.00963019597387015\\
12.7712125471966	0.00287432342354353\\
14.7712125471966	0.000793534195440608\\
16.7712125471966	0.000207225703239568\\
18.7712125471966	5.22730302626316e-05\\
20.7712125471966	1.2571657112385e-05\\
22.7712125471966	3.01293160911878e-06\\
24.7712125471966	8.26556459138781e-07\\
};

\addplot [color=mycolor1]
  table[row sep=crcr]{%
-2.21848749616356	0.652297880847661\\
-0.218487496163563	0.552860455817673\\
1.78151250383644	0.433900039984006\\
3.78151250383644	0.308366004532729\\
5.78151250383644	0.193980749233436\\
7.78151250383644	0.107471777096387\\
9.78151250383644	0.0526733235568591\\
11.7815125038364	0.0231819757365685\\
13.7815125038364	0.00934768697506999\\
15.7815125038364	0.0035164911345154\\
17.7815125038364	0.00127601653112918\\
19.7815125038364	0.000449220103986135\\
21.7815125038364	0.000156859085455273\\
23.7815125038364	5.36195173976803e-05\\
};
\addlegendentry{T/2 RRC}

\addplot [color=mycolor1, dashed, forget plot]
  table[row sep=crcr]{%
-0.457574905606752	0.698652538984406\\
1.54242509439325	0.563152738904438\\
3.54242509439325	0.397466613354658\\
5.54242509439325	0.233822237035062\\
7.54242509439325	0.110996533795494\\
9.54242509439325	0.0422535395280629\\
11.5424250943932	0.0131637914944674\\
13.5424250943932	0.0034888948140248\\
15.5424250943932	0.000808398880149313\\
17.5424250943932	0.000168777496333822\\
19.5424250943932	3.07692307692308e-05\\
21.5424250943932	5.54592720970537e-06\\
23.5424250943932	9.59872017064391e-07\\
25.5424250943932	1.06652446340488e-07\\
};
\addlegendentry{data2}

\end{axis}
\end{tikzpicture}%
	\caption{Comparison of conventional ISI-free PAM transmission and $\frac{T}{2}$ pulse shaping. Transmission is over $n = 2$ (solid) and $n = 3$ (dashed) antennas with one bit per real dimension ($M = 16$ and $M = 64$). Simulations were performed over several thousand Rayleigh fading channels and averaged.}
	\label{fig:rrcT2}
\end{figure}
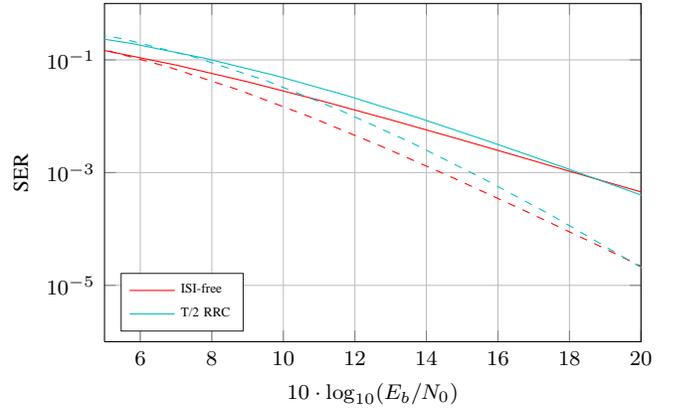

It is obvious that this method has a huge disadvantage due to occupying twice the spectrum. The reason why we nevertheless include it in this comparison is that $\frac{T}{2}$-pulse shaping increases the slope of the error curve such that in the medium- to high-SNR regime it might still be a valid alternative given the largely reduced PASPR compared to conventional PAM. Results for this pulse shaping method are plotted in Fig.~\ref{fig:rrcT2}.

The increased slope can be explained by the linear transformation of the hypersphere induced by $\vec H$: At high SNRs, symbol errors will usually occur because the noise moves the data symbol into the decision region of a neighboring symbol. Symbol errors to the opposite side of the hypersphere occur only rarely, because such points are farthest apart. If the receiver constellation $\vec{H}\mathcal{A} = \left\{\vec{Hx} \; | \; \vec{x} \in \mathcal{A} \right\}$ is distorted enough, such errors may be much more likely because the distance between opposing points might be drastically reduced. In the case of $\frac{T}{2}$-pulse shaping, not only the distance between constellation points, but also the distance between interpolation points affects the performance of the system. The spherical interpolation thus corresponds to a nonlinear convolutional code.  For data points on opposing sides of the hypersphere, $\frac{T}{2}$-pulse shaping generates interpolation points which are usually far away from each other. 
This increases the total minimum distance and thus the performance of the transmission. The magnitude of this effect is dependent on $\vec{H}$. A good measure for this effect is the ratio $\frac{\sigma_{\mathrm{SVD}, \mathrm{max}}}{\sigma_{\mathrm{SVD}, \mathrm{min}}}$ with $\sigma_{\mathrm{SVD}, \mathrm{max}}$ and $\sigma_{\mathrm{SVD}, \mathrm{min}}$ being the maximum and minimum singular values of the real representation of $\vec{H}$, respectively\footnote{Every complex-valued model $\vec{y} = \vec{Hx} + \vec n $ of dimension $n$ can be transformed into an equivalent real-valued model of dimension $2n$, see e.g.~\cite{realChannelModel}.}. The two extreme cases would be $\frac{\sigma_{\mathrm{SVD}, \mathrm{max}}}{\sigma_{\mathrm{SVD}, \mathrm{min}}} = 1$ in which $\vec{H}\mathcal{A}$ would still be a hypersphere (possibly with a different radius) and $\frac{\sigma_{\mathrm{SVD}, \mathrm{max}}}{\sigma_{\mathrm{SVD}, \mathrm{min}}} \rightarrow \infty$. In the latter case, the $n$-dimensional hypersphere would be compressed down to fewer dimensions, effectively reducing the distance between opposing points and possibly making them nearest neighbors.

\subsection{Spherical Interpolation}

The main problem of $\frac{T}{2}$-pulse shaping is the large bandwidth due to the use of pulse shaping filters at higher frequency. In order to mitigate the problem, we combine SI with a conventional pulse shaping filter being $\sqrt\mathrm{Nyquist}$ with respect to $T$.

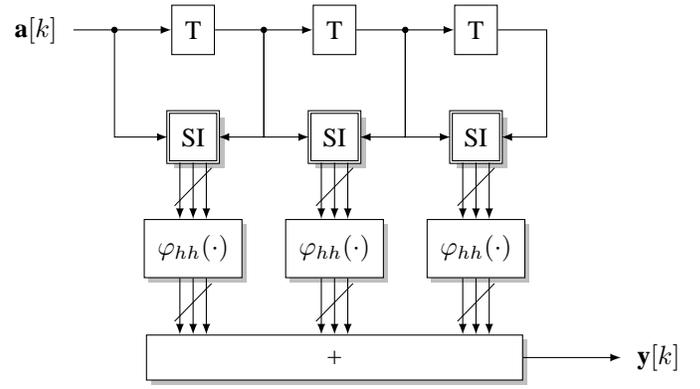
\begin{figure}[t]
	\setlength\figureheight{4.5cm}
	\setlength\figurewidth{7.5cm} 
	\begin{tikzpicture}[inner sep=.6em,x=1.3cm,y=1.5cm]
\node at (0,0) (in) {$\vec{a}[k]$};
\node[delayEle,right=1, at=(in.east)] (D1) {T};
\node[delayEle,right=1, at=(D1.east)] (D2) {T};
\node[delayEle,right=1, at=(D2.east)] (D3) {T};
\node at ($(in)!.5!(D1)$) (p1) {};
\node at ($(D1)!.5!(D2)$) (p2) {};
\node at ($(D2)!.5!(D3)$) (p3) {};
\draw[fill=black,->] ( $(in)!.5!(D1)$ ) circle(1pt);
\draw[fill=black,->] ( $(D1)!.5!(D2)$ ) circle(1pt);
\draw[fill=black,->] ( $(D2)!.5!(D3)$ ) circle(1pt);
\draw[->] (in.east) -- (D1.west);
\draw[->] (D1.east) -- (D2.west);
\draw[->] (D2.east) -- (D3.west);

\node[sysnonlinear, below = 0.5, at=(D1.south)] (SI1) {SI};
\node[sysnonlinear, below = 0.5, at=(D2.south)] (SI2) {SI};
\node[sysnonlinear, below = 0.5, at=(D3.south)] (SI3) {SI};

\draw[->] (p1.center) |- (SI1.west);
\draw[->] (p2.center) |- (SI1.east);
\draw[->] (p2.center) |- (SI2.west);
\draw[->] (p3.center) |- (SI2.east);
\draw[->] (p3.center) |- (SI3.west);
\draw[->] (D3.east) -- +(0.5, 0) |- (SI3.east);

\node[syslinear, below = 0.5, at=(SI1.south)] (rho1) {$\varphi_{hh}(\cdot)$};
\node[syslinear, below = 0.5, at=(SI2.south)] (rho2) {$\varphi_{hh}(\cdot)$};
\node[syslinear, below = 0.5, at=(SI3.south)] (rho3) {$\varphi_{hh}(\cdot)$};

\draw[->] ($(SI1.south)-(5pt,0)$) -- ($(rho1.north) -(5pt, 0)$);
\draw[->] (SI1.south) -- (rho1.north);
\draw[->] ($(SI1.south)+(5pt,0)$) -- ($(rho1.north) +(5pt, 0)$);
\draw[very thin] (SI1.south) ++(7pt, -2pt) -- ($(rho1.north) +(-7pt, 5pt)$);

\draw[->] ($(SI2.south)-(5pt,0)$) -- ($(rho2.north) -(5pt, 0)$);
\draw[->] (SI2.south) -- (rho2.north);
\draw[->] ($(SI2.south)+(5pt,0)$) -- ($(rho2.north) +(5pt, 0)$);
\draw[very thin] (SI2.south) ++(7pt, -2pt) -- ($(rho2.north) +(-7pt, 5pt)$);

\draw[->] ($(SI3.south)-(5pt,0)$) -- ($(rho3.north) -(5pt, 0)$);
\draw[->] (SI3.south) -- (rho3.north);
\draw[->] ($(SI3.south)+(5pt,0)$) -- ($(rho3.north) +(5pt, 0)$);
\draw[very thin] (SI3.south) ++(7pt, -2pt) -- ($(rho3.north) +(-7pt, 5pt)$);

\node[syslinear, minimum width = 5cm,  below = 0.5, at=(rho2.south)] (sum) {+};
\node[below = 0.5, at=(rho1.south)] (sum1) {};
\node[below = 0.5, at=(rho2.south)] (sum2) {};
\node[below = 0.5, at=(rho3.south)] (sum3) {};

\draw[->] ($(rho1.south)-(5pt,0)$) -- ($(sum1.north) -(5pt, 0)$);
\draw[->] (rho1.south) -- (sum1.north);
\draw[->] ($(rho1.south)+(5pt,0)$) -- ($(sum1.north) +(5pt, 0)$);
\draw[very thin] (rho1.south) ++(7pt, -2pt) -- ($(sum1.north) +(-7pt, 5pt)$);

\draw[->] ($(rho2.south)-(5pt,0)$) -- ($(sum2.north) -(5pt, 0)$);
\draw[->] (rho2.south) -- (sum2.north);
\draw[->] ($(rho2.south)+(5pt,0)$) -- ($(sum2.north) +(5pt, 0)$);
\draw[very thin] (rho2.south) ++(7pt, -2pt) -- ($(sum2.north) +(-7pt, 5pt)$);

\draw[->] ($(rho3.south)-(5pt,0)$) -- ($(sum3.north) -(5pt, 0)$);
\draw[->] (rho3.south) -- (sum3.north);
\draw[->] ($(rho3.south)+(5pt,0)$) -- ($(sum3.north) +(5pt, 0)$);
\draw[very thin] (rho3.south) ++(7pt, -2pt) -- ($(sum3.north) +(-7pt, 5pt)$);

\node[right = 1, at=(sum.east)] (output) {$\vec{y}[k]$};
\draw[->] (sum.east) -- (output.west);
\end{tikzpicture}
	\caption{System used to model the ISI produced by spherical interpolation transmission. An SI block calculates $f_\mathrm{IP}$ vectors and $\varphi_{hh}(\cdot)$ weighs them with the autocorrelation of the pulse shaping filter. Thus each block processes all interpolation vectors within one symbol period. This model omits the channel matrix $\vec{H}$ and noise $\vec{n}$. This example employs $\nu = 3$ memory elements.}
	\label{fig:FIR_SI}
\end{figure}

This method is characterized by the interpolation frequency $f_\mathrm{IP} \in \mathbb{N}$: In each symbol interval, the original data point as well as $f_\mathrm{IP} - 1$ interpolation points are transmitted. $f_\mathrm{IP} = 1$ corresponds to conventional PAM without SI. The resulting output signal of the transmitter is

\begin{align}
	\vec s (t) &= \sum_{k = -\infty}^{\infty} \left( \vphantom{ \sum_{l = 1}^{f_\mathrm{IP}-1}}  \vec x [k] \, h(t - kT) \right. \nonumber \\
			&\left. + \sum_{l = 1}^{f_\mathrm{IP}-1} \vec{SI}\left( \vec{x}[k], \vec{x}[k+1], \frac{l}{f_\mathrm{IP}}  \right) \, h\left(t - \left(k + \frac{l}{f_\mathrm{IP}} \right) T\right) \right).
\end{align}
At the receiver, matched filtering with $h^\ast(-t)$ is performed followed by $T$-spaced sampling. This is slightly suboptimal, but simplifies the receiver structure greatly. 
Introducing the autocorrelation \begin{equation}
\varphi_{hh}(\tau) = \int_{-\infty}^{\infty} h(t+\tau) h^\star(t) \mathrm{d}t,
\end{equation} the received discrete-time signal is
\begin{align}
	\vec{y}[k] &= \vec{y}(kT) = \vec{H} \left( \sum_{\bar{k} = -\infty}^{\infty} \vec{x}[\bar{k}] \varphi_{hh}(kT - \bar{k}T)  \right. \nonumber \\ 
			   &+ \sum_{l = 1}^{f_\mathrm{IP}-1} \vec{SI}\left( \vec{x}[\bar{k}], \vec{x}[\bar{k}+1], \frac{l}{f_\mathrm{IP}} \right) \nonumber \\
			   &\cdot  \left. \varphi_{hh}\left( kT -  \left(\bar{k} + \frac{l}{f_\mathrm{IP}}\right) T\right)  \right) + \vec{n}[k].
\end{align}
By using a $\sqrt{\mathrm{Nyquist}}$-pulse and $T$-spaced sampling, the direct influence of adjacent data symbols may be suppressed and the resulting noise at the receiver is white, but the influence of the interpolation values inserted at rate $T_\mathrm{IP} = T / f_\mathrm{IP}$ remains. Thus ISI-equalization in form of MLSE via the VA has to be performed at the receiver. Fig.~\ref{fig:FIR_SI} shows the system model used at the receiver to estimate the data sequence. The VA has to consider all $f_\mathrm{IP}$ vectors, which were transmitted during one symbol interval, to calculate a branch metric. Since $T$-spaced sampling is used, it is vital to use the contribution from the interpolation values, otherwise serious ISI would be unprocessed and its energy would be wasted. 

For SI pulse shaping, the choice of the filter does have an influence on the error probability, because it determines the shape of $\varphi_{hh}$. For the following results, we used a root-raised cosine (RRC) pulse shaping filter with roll-off factor $\beta = 0.25$.  Additionally, the choice of $f_\mathrm{IP}$ provides a trade-off between receiver complexity and smoothness of the output signal (which in term improves PASPR and bandwidth, see Sec.~\ref{sec:paspr}). The results in Figs.~\ref{fig:paspr} and \ref{fig:spectra} were generated using $f_\mathrm{IP} = 4$. Increasing $f_\mathrm{IP}$ to 16 showed a 0.15 dB improvement in PASPR, whereas the error probability is unaffected by increasing $f_\mathrm{IP}$.

In order to achieve full ML detection, all coefficients of the impulse response have to be considered and MLSE can be performed using the VA. The system model for $\nu = 3$ memory elements is depicted in Fig.~\ref{fig:FIR_SI}: Two adjacent symbols in time generate SI data, which is weighted according to $\varphi_{hh}$ and summed up. Given the size of PSKH constellations, full ML detection is computationally impossible. Thus we need to apply various complexity reduction methods, which are described in the next subsection.


\subsection{Complexity Reduction Techniques}
\begin{figure}[t]
	\setlength\figureheight{4.5cm}
	\setlength\figurewidth{7.5cm} 
%
%
\definecolor{mycolor1}{rgb}{0.00000,0.74902,0.74902}%
\begin{tikzpicture}

\begin{axis}[%
width=0.951\figurewidth,
height=\figureheight,
at={(0\figurewidth,0\figureheight)},
scale only axis,
xmin=5,
xmax=20,
xlabel style={font=\color{white!15!black}},
xlabel={$10\cdot \log_{10}(E_b/N_0)$},
ymode=log,
ymin=1e-06,
ymax=1,
yminorticks=true,
ylabel style={font=\color{white!15!black}},
ylabel={SER},
axis background/.style={fill=white},
xmajorgrids,
ymajorgrids,
yminorgrids,
legend style={at={(0.03,0.03)}, anchor=south west, legend cell align=left, align=left, draw=white!15!black},
ylabel style={font=\footnotesize},xlabel style={font=\footnotesize},legend style={font=\tiny},
]
\addplot [color=red]
  table[row sep=crcr]{%
-4.98970004336019	0.669436741767764\\
-2.98970004336019	0.579219370750567\\
-0.989700043360188	0.470592800959872\\
1.01029995663981	0.352190814558059\\
3.01029995663981	0.239099186775097\\
5.01029995663981	0.146112065057992\\
7.01029995663981	0.0806206505799227\\
9.01029995663981	0.0406765497933609\\
11.0102999566398	0.0191661245167311\\
13.0102999566398	0.0086091454472737\\
15.0102999566398	0.00376097853619517\\
17.0102999566398	0.00161137181709105\\
19.0102999566398	0.000689028129582722\\
21.0102999566398	0.000297733635515265\\
23.0102999566398	0.000127982935608586\\
};
\addlegendentry{ISI-free};
\addplot [color=red, dashed, forget plot]
  table[row sep=crcr]{%
-3.22878745280338	0.721007958938808\\
-1.22878745280338	0.601410038661512\\
0.771212547196624	0.453156312491668\\
2.77121254719662	0.297072310358619\\
4.77121254719662	0.164149073456872\\
6.77121254719662	0.0753689374750033\\
8.77121254719662	0.0289929742700973\\
10.7712125471966	0.00963019597387015\\
12.7712125471966	0.00287432342354353\\
14.7712125471966	0.000793534195440608\\
16.7712125471966	0.000207225703239568\\
18.7712125471966	5.22730302626316e-05\\
20.7712125471966	1.2571657112385e-05\\
22.7712125471966	3.01293160911878e-06\\
24.7712125471966	8.26556459138781e-07\\
};
\addplot [color=mycolor1]
  table[row sep=crcr]{%
-4.98970004336019	0.711929614077185\\
-2.98970004336019	0.634589282143571\\
-0.989700043360188	0.535029794041192\\
1.01029995663981	0.416241351729654\\
3.01029995663981	0.291914817036593\\
5.01029995663981	0.182177964407119\\
7.01029995663981	0.101102179564087\\
9.01029995663981	0.05049450109978\\
11.0102999566398	0.023249744478514\\
13.0102999566398	0.01011825089988\\
15.0102999566398	0.00424076789761365\\
17.0102999566398	0.00177522996933742\\
19.0102999566398	0.000758165577922944\\
};
\addlegendentry{SI RRC, $\beta = 0.25$};
\addplot [color=mycolor1, dashed, forget plot]
  table[row sep=crcr]{%
0.771212547196624	0.522643691952184\\
2.77121254719662	0.357188997022619\\
4.77121254719662	0.203236857307914\\
6.77121254719662	0.0946634670932765\\
8.77121254719662	0.0366262276140959\\
10.7712125471966	0.01221272719193\\
12.7712125471966	0.003667866506688\\
14.7712125471966	0.00103661733991023\\
16.7712125471966	0.000289428076256499\\
18.7712125471966	8.44776252055281e-05\\
20.7712125471966	2.52855174865573e-05\\
};
\end{axis}
\end{tikzpicture}%
	\caption{Comparison of ISI-free transmission and SI pulse shaping with $f_\mathrm{IP} = 4$. Transmission is over $n = 2$ (solid) and $n = 3$ (dashed) antennas with one bit per real dimension ($M = 16$ and $M = 64$). Detection was performed using DDFSE employing $\nu = 2$ memory elements for $n = 3$ antennas and $\nu = 3$ memory elements for $n = 2$ antennas. Simulations are averaged over several thousand Rayleigh fading channels.}
	\label{fig:SI}
\end{figure}
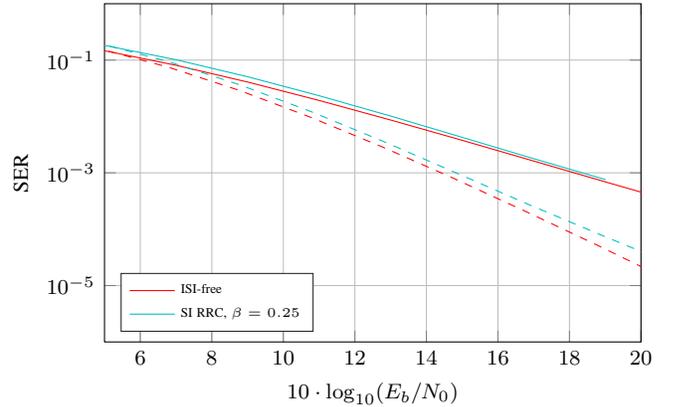

In order to make a detector computationally feasible, we only use those intervals of $\varphi_{hh}$ which have the highest energy and use it for sequence estimation in the VA. Prior taps are considered as noise and remaining taps at the end are equalized using DFE, which makes the overall scheme a Delayed Decision-Feedback Sequence Estimation (DDFSE)~\cite{DDFSE}. Fig.~\ref{fig:SI} shows numerical results for the spherical interpolation shaping using $f_\mathrm{IP} = 4$. Detection was performed using DDFSE employing $\nu = 3$ memory elements for the 16-ary constellation and $\nu = 2$ memory elements for the 64-ary constellation (corresponding to 4096 states in both cases).

For DDFSE, usually a prefilter is applied to make the overall impulse response minimum-phase, see e.g.~\cite{MinPhasePrefilter}. In this case, we have to deal with a number of problems: First of all, SI is  a nonlinear operation. Secondly, the overall impulse response has a large linear phase portion. Thirdly, for a given interval width, the minimum phase response captures less of the total energy than the overall impulse response: The original impulse response has a large fraction of its total energy concentrated around the center, whereas the minimum phase part spreads its energy over a wider interval in the beginning of the impulse. Thus, more memory elements in the VA are required to capture the same amount of energy. This is computationally not feasible. Fourthly, given the total length of $\varphi_{hh}$, numerical inaccuracies might occur when calculating the prefilter. It is thus advantageous to use the original filter instead of applying any prefilter to make the overall filter minimum phase and simply treat the influence of the first taps as noise.

The power efficiency of the SI pulse compared to conventional PAM employing a RRC filter depends on the roll-off factor and the number of states in the VA: For a fixed number of states in the receiver, increasing the roll-off factor $\beta$ of the RRC filter improves power efficiency, because more of the energy of $\varphi_{hh}(t)$ is concentrated around the center. This is in contrast to conventional PAM, where power efficiency is unaffected by the roll-off factor. Our analyses show that for $\beta = 0.1$, a loss of approx. 1.1 dB occurs at a target symbol error rate of $10^{-4}$. This loss shrinks quickly with an increasing roll-off factor: For $\beta = 0.25$, the gap is closed. Increasing the number of memory elements also reduces the loss, because the amount of energy used for sequence estimation is increased. The feasibility of this is restricted by computational complexity. Therefore, we now discuss how the gap between SI RRC and conventional RRC can be closed with reduced computational complexity.

It is usually sufficient to use only 2 or 3 delay elements to capture almost all energy of the pulse. The remaining energy at the end of the filter can be equalized by means of DDFSE. But since every delay element is $M$-ary, the number of states can become infeasible even for such a small number of elements. We thus compare system performance and complexity when using two different complexity reduction techniques: The well-known Reduced State Sequence Estimation (RSSE)~\cite{RSSE} and a newly proposed iterative application of the VA.

For RSSE, we use a Viterbi algorithm with $\nu = 2$ memory elements and generate hyperstates by using hypersymbols in the second delay element only. Combining different input symbols into a hypersymbol in the first element leads to large performance degradation due to two effects: Hyperstates are calculated in advance based on the original constellation $\mathcal{A}$. We did this by numerically optimizing the minimum distance within each hyperstate~\cite{art:spinnler}. The effective constellation at the receiver, however, is $\vec{H}\mathcal{A}$ which might have a drastically different distance profile than the constellation with originally optimal hyperstates. The other negative impact is the fact that both the impulse response $\varphi_{hh}(t)$ as well as its minimum-phase component do not have monotonously decreasing values. The decision in the first delay element is thus based on only a small fraction of the total pulse energy. 

Our second approach to reduce the complexity is to apply the Viterbi algorithm iteratively. This works well if the performance gap between the use of $\nu$ and $\nu + 1$ memory elements is not too large. The idea behind it is that if each error pattern is a neighboring symbol of the correct signal point, it is sufficient to consider these neighboring symbols in future steps. This is a valid assumption if the SNR is sufficiently high. Our algorithm works as follows:
\begin{enumerate}
	\item Initialize $\nu = 1$.
	\item Run the Viterbi algorithm with one memory element.
	\item For each estimate $\hat{\vec{x}}[k]$, find the $n_\mathrm{NB}$ nearest neighboring points.
	\item Set $\nu = \nu + 1$.
	\item Run the Viterbi algorithm with $\nu$ memory elements, only allowing $n_\mathrm{NB} + 1$ points in each time step.
	\item If $\nu = \nu_\mathrm{max}$ finish, otherwise go back to 3.
\end{enumerate}

The neighboring points can be calculated in advance and stored in a table. This works best if the neighboring points are taken from $\vec{H}\mathcal{A}$, but a reasonable performance can also be achieved if they are taken directly from $\mathcal{A}$, which reduces the overhead to recalculate them every time $\vec{H}$ changes.
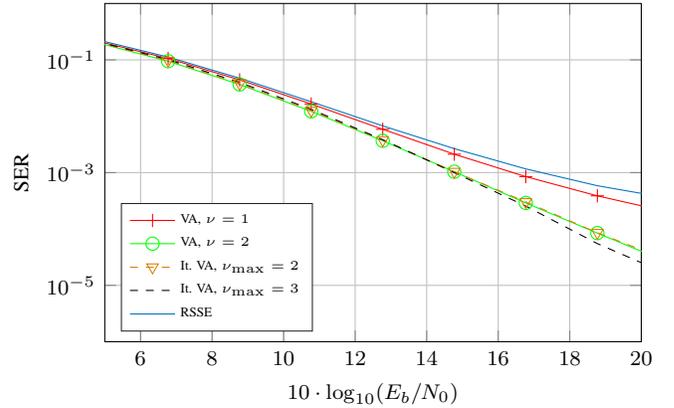
\begin{figure}[t]
	\setlength\figureheight{4.5cm}
	\setlength\figurewidth{7.5cm} 
%
%
\definecolor{mycolor1}{rgb}{0.87059,0.49020,0.00000}%
\definecolor{mycolor2}{rgb}{0.00000,0.44706,0.74118}%
\begin{tikzpicture}

\begin{axis}[%
width=0.951\figurewidth,
height=\figureheight,
at={(0\figurewidth,0\figureheight)},
scale only axis,
xmin=5,
xmax=20,
xlabel style={font=\color{white!15!black}},
xlabel={$10\cdot \log_{10}(E_b/N_0)$},
ymode=log,
ymin=1e-06,
ymax=1,
yminorticks=true,
ylabel style={font=\color{white!15!black}},
ylabel={SER},
axis background/.style={fill=white},
xmajorgrids,
ymajorgrids,
yminorgrids,
legend style={at={(0.03,0.03)}, anchor=south west, legend cell align=left, align=left, draw=white!15!black},
ylabel style={font=\footnotesize},xlabel style={font=\footnotesize},legend style={font=\tiny},
]
\addplot [color=red, mark=+, mark options={solid, red}]
  table[row sep=crcr]{%
0.771212547196624	0.532491838487973\\
2.77121254719662	0.370231958762887\\
4.77121254719662	0.216899341351661\\
6.77121254719662	0.105889747995418\\
8.77121254719662	0.0441179839633448\\
10.7712125471966	0.0164873997709049\\
12.7712125471966	0.00586970217640321\\
14.7712125471966	0.00211898625429553\\
16.7712125471966	0.00084758018327606\\
18.7712125471966	0.000387099083619702\\
20.7712125471966	0.00019967067583047\\
};
\addlegendentry{VA, $\nu = 1$}

\addplot [color=green, mark=o, mark options={solid, green}]
  table[row sep=crcr]{%
0.771212547196624	0.522643691952184\\
2.77121254719662	0.357188997022619\\
4.77121254719662	0.203236857307914\\
6.77121254719662	0.0946634670932765\\
8.77121254719662	0.0366262276140959\\
10.7712125471966	0.01221272719193\\
12.7712125471966	0.003667866506688\\
14.7712125471966	0.00103661733991023\\
16.7712125471966	0.000289428076256499\\
18.7712125471966	8.44776252055281e-05\\
20.7712125471966	2.52855174865573e-05\\
};
\addlegendentry{VA, $\nu = 2$}

\addplot [color=mycolor1, dashed, mark=triangle, mark options={solid, rotate=180, mycolor1}]
  table[row sep=crcr]{%
0.771212547196624	0.527906145624582\\
2.77121254719662	0.362677521710087\\
4.77121254719662	0.207539245156981\\
6.77121254719662	0.097063460253841\\
8.77121254719662	0.0377556780227121\\
10.7712125471966	0.0125492651970608\\
12.7712125471966	0.00377638610554442\\
14.7712125471966	0.00103957915831663\\
16.7712125471966	0.000302271209084836\\
18.7712125471966	8.58383433533734e-05\\
20.7712125471966	2.75551102204409e-05\\
};
\addlegendentry{It. VA, $\nu_\mathrm{max} = 2$}

\addplot [color=black, dashed]
  table[row sep=crcr]{%
0.771212547196624	0.532838\\
2.77121254719662	0.3683285\\
4.77121254719662	0.212725\\
6.77121254719662	0.101013\\
8.77121254719662	0.0398855\\
10.7712125471966	0.0132535\\
12.7712125471966	0.003838\\
14.7712125471966	0.000998\\
16.7712125471966	0.0002545\\
18.7712125471966	5.5e-05\\
20.7712125471966	1.55e-05\\
};
\addlegendentry{It. VA, $\nu_\mathrm{max} = 3$}

\addplot [color=mycolor2]
  table[row sep=crcr]{%
0.771212547196624	0.551855428914217\\
2.77121254719662	0.387998400319936\\
4.77121254719662	0.2290549890022\\
6.77121254719662	0.112175164967007\\
8.77121254719662	0.0473063387322535\\
10.7712125471966	0.0180789842031594\\
12.7712125471966	0.00674151311717374\\
14.7712125471966	0.00265919154406513\\
16.7712125471966	0.00116231014616959\\
18.7712125471966	0.000585202113983717\\
20.7712125471966	0.000351806884730753\\
};
\addlegendentry{RSSE}

\end{axis}
\end{tikzpicture}%
	\caption{Comparison of SI pulse shaping for $n = 3$ antennas with a RRC pulse shape with $\beta = 0.25$. RSSE used a quaternary second delay element and $n_\mathrm{NB} = 4$ for all iterative variants. Simulations were performed over several thousand Rayleigh fading channels and averaged.}
	\label{fig:complRed}
\end{figure}

Fig.~\ref{fig:complRed} shows the performance of a 64-ary alphabet transmitted via SI signaling with a RRC pulse with $\beta = 0.25$ employing $n = 3$ antennas. The VA curve using two memory elements ($\nu = 2$) is the same as in Fig.~\ref{fig:SI} and is our baseline. As a measure of complexity we count the number of trellis branches in each time step
\begin{equation}
	 \Xi = \begin{cases} 
				  M^{\nu+1},\; \; &\text{Standard VA} \\
				  M \cdot \prod_{i = 1}^{\nu} M_i, \; \; &\text{VA, RSSE} \\
				  M^2 + \sum_{i=2}^{\nu_\mathrm{max}} (n_\mathrm{NB, i} + 1)^{\nu_i+1}, \; \; &\text{Iterative VA}.
			\end{cases}
\end{equation}
In this term, $M_i$ is the number of possible values in the $i$-th delay element if RSSE is used (the number of hypersymbols) and $\nu_\mathrm{i}$ is the number of memory elements in the $i$-th iteration of the iterative VA. Table~\ref{tbl:complexity} shows the complexity for the algorithms used to create Fig.~\ref{fig:complRed}. The computational complexity for a VA with $\nu = 2$ is already impractical. The iterative VA, however, provides the same performance as the VA with $\nu = 2$ with only minor complexity increase compared to the VA with $\nu = 1$. Increasing the number of iterations by one allows to improve the power efficiency (approx. 0.5 dB at $\mathrm{SER} = 10^{-4}$) such that the iterative VA outperforms the VA with two delay elements.
The exact results for the iterative VA depend on the shape of the overall impulse response which changes with the roll-off factor. For practical values $\beta > 0.2$, we found the differences to be only marginal.

\begin{table}[!t]
	\renewcommand{\arraystretch}{1.3}
	\caption{Complexity Comparison of \newline SI Demodulation (M = 64) for Fig.~\ref{fig:complRed}}
	\label{tbl:complexity}
	\centering
	\begin{tabular}{l||l}
		\hline
		Algorithm & $\Xi$ \\
		\hline\hline
		Viterbi, $\nu = 1$ 	& 4096  \\
		Viterbi, $\nu = 2$ 	& 262144  \\
		Viterbi, RSSE 		& 16384 \\
		It. Viterbi, $\nu_\mathrm{max} = 2$ &  4221 \\
		It. Viterbi, $\nu_\mathrm{max} = 3$ &  4846 \\
		\hline\hline
	\end{tabular}
\end{table}


\section{PASPR, Spectrum and Bandwidth Efficiency}
\label{sec:paspr}

\begin{figure}[t]
	\setlength\figureheight{4.5cm}
	\setlength\figurewidth{7.5cm} 
	\input{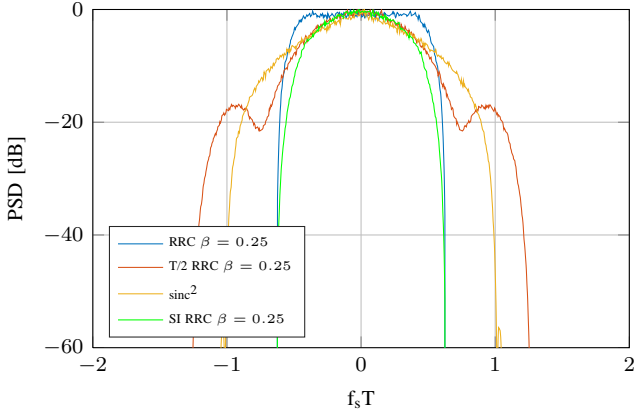}
	\caption{Occupied spectra of different pulse shaping methods. The spectra are calculated for a $n = 4$ antenna system.}
	\label{fig:spectra}
\end{figure}

\begin{figure}[t]
	\setlength\figureheight{4.5cm}
	\setlength\figurewidth{7.5cm} 
%
%
\definecolor{mycolor1}{rgb}{0.00000,0.44700,0.74100}%
\definecolor{mycolor2}{rgb}{0.85000,0.32500,0.09800}%
\definecolor{mycolor3}{rgb}{0.92900,0.69400,0.12500}%
\definecolor{mycolor4}{rgb}{0.49400,0.18400,0.55600}%
\definecolor{mycolor5}{rgb}{0.46600,0.67400,0.18800}%
\definecolor{mycolor6}{rgb}{0.30100,0.74500,0.93300}%
\begin{tikzpicture}

\begin{axis}[%
width=0.951\figurewidth,
height=\figureheight,
at={(0\figurewidth,0\figureheight)},
scale only axis,
xmin=1,
xmax=7,
xlabel={Number of Antennas},
xmajorgrids,
ymin=0,
ymax=6,
ylabel={PASPR [dB]},
ymajorgrids,
axis background/.style={fill=white},
legend style={legend cell align=left,align=left,draw=white!15!black},
ylabel style={font=\footnotesize},xlabel style={font=\footnotesize},legend style={font=\tiny},
]
\addplot [color=mycolor1,solid]
  table[row sep=crcr]{%
1	5.03\\
2	4.71\\
3	4.36\\
4	4.20\\
5	3.74\\
6	3.48\\
7	3.37\\
};
\addlegendentry{$\text{RRC }\beta = 0.25$};

\addplot [color=mycolor2,solid]
  table[row sep=crcr]{%
1	1.75440722120997\\
2	1.58795672402732\\
3	1.37160770566896\\
4	1.16419361166331\\
5	1.03689122285632\\
6	0.917376114543668\\
7	0.857451244296939\\
};
\addlegendentry{$\text{T/2 RRC }\beta = 0.25$};

\addplot [color=mycolor3,solid]
  table[row sep=crcr]{%
1	1.76121282000213\\
2	1.76106951614112\\
3	1.76101515516577\\
4	1.76074585878946\\
5	1.7609666251559\\
6	1.76101627280874\\
7	1.7607222573299\\
};
\addlegendentry{$\text{sinc}^\text{2}$};

%

\addplot [color=green,solid]
  table[row sep=crcr]{%
1	3.34427420914064\\
2	2.79321323245524\\
3	2.44032791572127\\
4	2.16771009642203\\
5	1.94309901312004\\
6	1.86178081085649\\
7	1.68426903923051\\
};
\addlegendentry{$\text{SI RRC }\beta = 0.25$};

\end{axis}
\end{tikzpicture}%
	\caption{Resulting PASPRs of different pulse shaping methods. A modulation rate of 1 bit per real dimension and an interpolation frequency $f_\mathrm{IP} = 4$ for SI RRC were used.}
	\label{fig:paspr}
\end{figure}
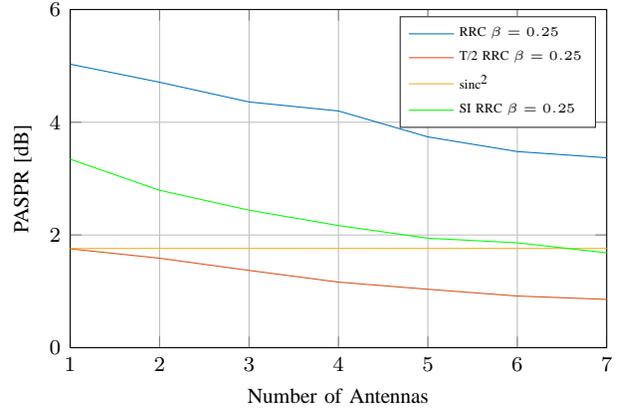

In the previous sections, we introduced several methods to reduce the PASPR of a signal and discussed their power efficiency. Some methods may have a negative impact on the bandwidth, but a wider spectrum may be tolerable, if the gain in PASPR is substantial. In this section we discuss how much PASPR reduction can be achieved and how the corresponding spectrum behaves.

Our baseline is a RRC pulse shaping filter with roll-off factor $\beta = 0.25$. The comparison of bandwidth and PASPR is given in Figs.~\ref{fig:spectra} and \ref{fig:paspr}, respectively. In these plots, RRC and $\mathrm{sinc}^2$ describe conventional pulse shaping (see Sec.~\ref{sec:sincSq} for $\mathrm{sinc}^2$ pulse shaping), T/2-RRC describes pulse shaping at twice the symbol rate and SI based pulse shaping is named SI RRC (see Sec.~\ref{sec:sphInt}).

A simple conclusion is that PAPSR reduction can be traded for bandwidth, i.e., the widest spectrum produces the lowest PASPR: T/2-RRC has the lowest PASPR, followed by $\mathrm{sinc}^2$ pulse shaping and SI RRC has the least PASPR reduction, but it also is the only method which does not widen the spectrum. One should also take into account the receiver complexity for these techniques: $\mathrm{sinc}^2$ requires almost no additional complexity compared to ISI-free PAM,  T/2-RRC and SI RRC require a sequence estimation to achieve a reasonable power efficiency. For all RRC based methods, the well-known trade-off between bandwidth and roll-off still holds. Additionally, increasing $\beta$ also improves the PASPR slightly.


\begin{figure*}[!t]
	\centering
	\hspace*{-1cm}
	\begin{tabular}{ll}
		\captionsetup[subfloat]{captionskip=0pt, nearskip=0.2cm, margin=0.5cm, justification=centering}
		\subfloat[Plot over PASPR]{
			\setlength\figureheight{4.5cm}
			\setlength\figurewidth{7.5cm} 		
%
%
\definecolor{mycolor1}{rgb}{0.00000,0.44700,0.74100}%
\definecolor{mycolor2}{rgb}{0.85000,0.32500,0.09800}%
\definecolor{mycolor3}{rgb}{0.92900,0.69400,0.12500}%
\definecolor{mycolor4}{rgb}{0.49400,0.18400,0.55600}%
\definecolor{mycolor5}{rgb}{0.46600,0.67400,0.18800}%
\definecolor{mycolor6}{rgb}{0.30100,0.74500,0.93300}%
\definecolor{mycolor7}{rgb}{0.63500,0.07800,0.18400}%
\begin{tikzpicture}

\begin{axis}[%
width=0.951\figurewidth,
height=\figureheight,
at={(0\figurewidth,0\figureheight)},
scale only axis,
xmin=1,
xmax=7,
xlabel={PASPR [dB]},
xmajorgrids,
ymin=2,
ymax=6.5,
ylabel={Spectral Efficiency [bit/s/Hz]},
ymajorgrids,
axis background/.style={fill=white},
legend style={at={(0.97,0.03)},anchor=south east,legend cell align=left,align=left,draw=white!15!black},
ylabel style={font=\footnotesize},xlabel style={font=\footnotesize},legend style={font=\tiny},
]
\addplot [color=mycolor1,only marks,mark=square,mark options={solid}]
  table[row sep=crcr]{%
6.15	6\\
4.35	4.8\\
3.15	4\\
};
\addlegendentry{RRC, varying $\beta$};

\addplot [color=mycolor2,only marks,mark=o,mark options={solid}]
  table[row sep=crcr]{%
1.76	3.75\\
1.76	3.333\\
1.76	3.15\\
1.76	3\\
};
\addlegendentry{$\mathrm{sinc}^2$};

\addplot [color=mycolor3,only marks,mark=square,mark options={solid}]
  table[row sep=crcr]{%
2.2	3.49\\
2.2	3.01\\
2.2	3\\
2.2	3\\
};
\addlegendentry{T/2-RRC, $\beta = 0$};

\addplot [color=mycolor4,only marks,mark=asterisk,mark options={solid}]
  table[row sep=crcr]{%
1.35	3.42\\
1.35	2.8\\
1.35	2.6\\
1.35	2.4\\
};
\addlegendentry{T/2-RRC, $\beta = 0.25$};

\addplot [color=mycolor5,only marks,mark=*,mark options={solid}]
  table[row sep=crcr]{%
1.7	3.125\\
1.7	2.54\\
1.7	2.11\\
1.7	2\\
};
\addlegendentry{T/2-RRC, $\beta = 0.5$};

\addplot [color=blue,only marks,mark=otimes,mark options={solid}]
table[row sep=crcr]{%
	3	6.25\\
	3	5.8252427184466\\
	3	5.66037735849057\\
	3	5.45454545454545\\
};
\addlegendentry{SI-RRC, $\beta = 0.1$};

\addplot [color=green,only marks,mark=diamond,mark options={solid}]
  table[row sep=crcr]{%
2.4	6.18\\
2.4	5.45\\
2.4	5.12\\
2.4	4.8\\
};
\addlegendentry{SI-RRC, $\beta = 0.25$};

\addplot [color=mycolor7,only marks,mark=triangle,mark options={solid,rotate=180}]
  table[row sep=crcr]{%
1.75	6\\
1.75	5.04\\
1.75	4.58\\
1.75	4\\
};
\addlegendentry{SI-RRC, $\beta = 0.5$};

\node (source) at (axis cs:2.1,4){$B_x$};
\node (destination) at (axis cs:2.1,6){};
\draw[<-](source)--(destination);
\node (b0) at (axis cs:5.7,6){\tiny $\beta = 0$};
\node (b25) at (axis cs:3.85,4.75){\tiny $\beta = 0.25$};
\node (b5) at (axis cs:3.6,4){\tiny $\beta = 0.5$};

\end{axis}
\end{tikzpicture}%
			\label{fig:spec_eff1}} &
		\captionsetup[subfloat]{captionskip=0pt, nearskip=0.2cm, margin=0.5cm, justification=centering}
		\subfloat[Plot over $E_{b, \mathrm{max}}/N_0 = E_b \cdot \mathrm{PASPR} / N_0$]{
			\setlength\figureheight{4.5cm}
			\setlength\figurewidth{7.5cm} 
			\hspace*{-.5cm}
%
%
\definecolor{mycolor1}{rgb}{0.00000,0.44700,0.74100}%
\definecolor{mycolor2}{rgb}{0.85000,0.32500,0.09800}%
\definecolor{mycolor3}{rgb}{0.92900,0.69400,0.12500}%
\definecolor{mycolor4}{rgb}{0.49400,0.18400,0.55600}%
\definecolor{mycolor5}{rgb}{0.46600,0.67400,0.18800}%
\definecolor{mycolor6}{rgb}{0.30100,0.74500,0.93300}%
\definecolor{mycolor7}{rgb}{0.63500,0.07800,0.18400}%
\begin{tikzpicture}

\begin{axis}[%
width=0.951\figurewidth,
height=\figureheight,
at={(0\figurewidth,0\figureheight)},
scale only axis,
xmin=19,
xmax=25,
xlabel={$10\cdot \log_{10}(E_{b, \mathrm{max}}/N_0)$},
xmajorgrids,
ymin=2,
ymax=6.5,
ylabel={Spectral Efficiency [bit/s/Hz]},
ymajorgrids,
axis background/.style={fill=white},
legend style={at={(0.97,0.03)},anchor=south east,legend cell align=left,align=left,draw=white!15!black},
ylabel style={font=\footnotesize},xlabel style={font=\footnotesize},legend style={font=\tiny},
]
\addplot [color=mycolor1,only marks,mark=square,mark options={solid}]
  table[row sep=crcr]{%
24.06	6\\
22.21	4.8\\
20.95	4\\
};
\addlegendentry{RRC, varying $\beta$};

\addplot [color=mycolor2,only marks,mark=o,mark options={solid}]
  table[row sep=crcr]{%
19.64	3.75\\
19.64	3.33333333333333\\
19.64	3.15789473684211\\
19.64	3\\
};
\addlegendentry{$\mathrm{sinc}^2$};

\addplot [color=mycolor3,only marks,mark=square,mark options={solid}]
  table[row sep=crcr]{%
20.35	3.48837209302326\\
20.35	3.06122448979592\\
20.35	3\\
20.35	3\\
};
\addlegendentry{T/2-RRC, $\beta = 0$};

\addplot [color=mycolor4,only marks,mark=asterisk,mark options={solid}]
  table[row sep=crcr]{%
19.45	3.42857142857143\\
19.45	2.80373831775701\\
19.45	2.60869565217391\\
19.45	2.4\\
};
\addlegendentry{T/2-RRC, $\beta = 0.25$};

\addplot [color=mycolor5,only marks,mark=*,mark options={solid}]
  table[row sep=crcr]{%
19.85	3.125\\
19.85	2.54237288135593\\
19.85	2.10526315789474\\
19.85	2\\
};
\addlegendentry{T/2-RRC, $\beta = 0.5$};

\addplot [color=blue,only marks,mark=otimes,mark options={solid}]
table[row sep=crcr]{%
22.1	6.25\\
22.1	5.8252427184466\\
22.1	5.66037735849057\\
22.1	5.45454545454545\\
};
\addlegendentry{SI-RRC, $\beta = 0.1$};

\addplot [color=green,only marks,mark=diamond,mark options={solid}]
  table[row sep=crcr]{%
20.3	6.18556701030928\\
20.3	5.45454545454545\\
20.3	5.12820512820513\\
20.3	4.8\\
};
\addlegendentry{SI-RRC, $\beta = 0.25$};

\addplot [color=mycolor7,only marks,mark=triangle,mark options={solid,rotate=180}]
  table[row sep=crcr]{%
19.5	6\\
19.5	5.04201680672269\\
19.5	4.58015267175572\\
19.5	4\\
};
\addlegendentry{SI-RRC, $\beta = 0.5$};

\node (source) at (axis cs:20,4){$B_x$};
\node (destination) at (axis cs:20,6){};
\draw[<-](source)--(destination);
\node (b0) at (axis cs:23.7,6){\tiny $\beta = 0$};
\node (b25) at (axis cs:21.7,4.75){\tiny $\beta = 0.25$};
\node (b5) at (axis cs:21.4,4){\tiny $\beta = 0.5$};
\end{axis}
\end{tikzpicture}%
			\label{fig:spec_eff2}} 
	\end{tabular}
	\caption{Spectral efficiencies for different pulse shaping averaged over many Rayleigh fading channels with $n = 3$ antennas and a constellation size of $M = 64$. For all methods, except RRC, we plot the spectral efficiency based on the bandwidth $B_x$ for a fraction $x$ of the total energy with $x \in \{99\%, 99.9\%, 99.99\%, 100\%\}$. A target symbol error rate of $\mathrm{SER} = 10^{-4}$ was used for the power bandwidth plane in Fig.~\ref{fig:spec_eff2}. Simulations were performed over several thousand Rayleigh fading channels and averaged.}
	\label{fig:spec_eff}
\end{figure*}
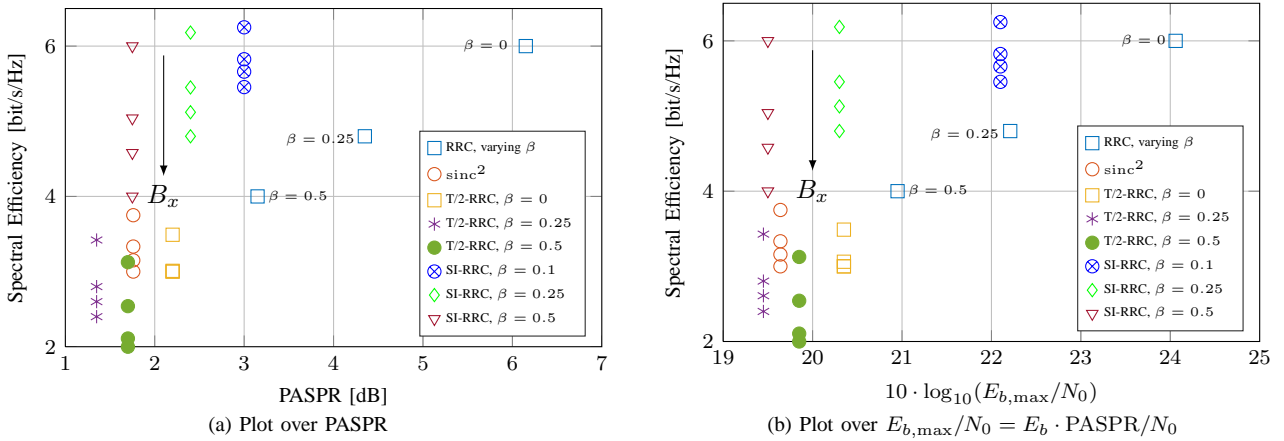

To summarize the results, we compare PASPR as well as power and spectral efficiencies of the methods presented in this paper. Since some pulse shaping methods have wide spectra, we also consider the bandwidth $B_x$ which includes a fraction $x$ of the total signal energy, e.g., $B_{99\%}$ is the bandwidth which holds $99\%$ of the total energy of a signal.

In Fig.~\ref{fig:spec_eff}, the spectral efficiencies are plotted for a transmission system employing $n = 3$ antennas and a constellation size of $M = 64$ over a Rayleigh fading channel. As a baseline, a RRC with $\beta = \{0, 0.25, 0.5\}$ is used. For all other pulse shaping methods, we plot the spectral efficiencies for $B_x$ with $x \in \{99\%, 99.9\%, 99.99\%, 100\%\}$. Fig.~\ref{fig:spec_eff1} plots the spectral efficiencies over the PASPR, whereas Fig.~\ref{fig:spec_eff2} uses the maximum energy per bit over the noise spectral density, i.e., $E_{b, \mathrm{max}}/N_0 = E_b \cdot \mathrm{PASPR} / N_0$. This takes both the maximum instantaneous power and the power efficiency for a given error rate into account and thus provides a fair comparison. These results are especially applicable for load-modulated transmitters.

The general result can be summarized as follows: All methods presented in this paper provide reasonable reduction of the PASPR. Some methods, however, do so at the cost of reduced spectral efficiency. SI pulse shaping is the only method to reduce the PASPR without sacrificing spectral efficiency. The cost to be paid in this case is a more complex receiver architecture.

In Fig.~\ref{fig:spec_eff2}, the power-bandwidth plane for a target symbol error rate of $\mathrm{SER} = 10^{-4}$ is shown. Because all losses in terms of power efficiency for a given symbol error rate are minor (if existing), especially SI RRC is superior to the conventional PAM. The gain due to the reduced PASPR generally outweighs the loss due to suboptimal detection of ISI if $\beta > 0.1$. Depending on the roll-off factor, the final gain in $E_{b, \mathrm{max}}/N_0$ is in the range of 1 to 2 dB. Substantial gain can also be achieved by T/2-RRC and $\mathrm{sinc}^2$ pulse shaping. Since these variants have almost no loss in power efficiency, they can realize their whole PASPR gain. The downside of them, again, is the reduced spectral efficiency.

\section{Conclusion}
\label{sec:conclusion}

PSKH is a novel modulation scheme for MIMO that is applicable in various scenarios: Because PSKH constellations are optimized over all antennas, both the constellation-constrained capacity as well as the error rate for a given power are improved compared to conventional PSK. This shows that PSKH is an interesting alternative even for MIMO systems which do not employ load modulation.

If PSKH is combined with load-modulation amplifiers, additional improvements are possible. The distribution on the hypersphere can be exploited to achieve a transmit signal with a low PASPR. By reducing the PASPR, amplifiers can be driven at a higher efficiency and thus the power loss is reduced. To achieve this, there is a trade-off between three degrees of freedom: Power efficiency, bandwidth efficiency and receiver complexity. It is possible to improve power efficiency at the cost of either bandwidth efficiency or receiver complexity. These results underline that load-modulation transmitters are a valid alternative for power-efficient communications of MIMO systems, which only employ a small number of antennas.

\ifCLASSOPTIONcaptionsoff
  \newpage
\fi



\bibliographystyle{IEEEtran}
\bibliography{IEEEabrv,references}
%



%




\end{document}